\documentclass[%
superscriptaddress,
preprint,
 amsmath,amssymb,
 aip,
apl,
]{revtex4-2}

\usepackage{graphicx}% Include figure files
\usepackage{dcolumn}% Align table columns on decimal point
\usepackage{bm}% bold math
\usepackage{gensymb}
\usepackage{hyperref}% add hypertext capabilities
\begin{document}

\title{Epitaxy of hexagonal ABO$_3$ quantum materials}

\author{Johanna Nordlander}
\affiliation{Department of Physics, Harvard University, USA
}%

\author{Margaret A. Anderson}
\affiliation{Department of Physics, Harvard University, USA
}%

\author{Charles M. Brooks}
\affiliation{Department of Physics, Harvard University, USA
}%

\author{Megan E. Holtz}
\affiliation{Department of Metallurgy and Materials Engineering, Colorado School of Mines, USA
}%

\author{Julia A. Mundy}
\email{mundy@fas.harvard.edu}
\affiliation{Department of Physics, Harvard University, USA
}%

\begin{abstract}
Hexagonal $AB$O$_3$ oxides ($A$, $B$ = cation) are a rich materials class for realizing novel quantum phenomena. Their hexagonal symmetry, oxygen trigonal bipyramid coordination and quasi-two dimensional layering give rise to properties distinct from those of the cubic $AB$O$_3$ perovskites. As bulk materials, most of the focus in this materials class has been on the rare earth manganites, $R$MnO$_3$ ($R$ = rare earth); these materials display coupled ferroelectricity and antiferromagnetic order. In this review, we focus on the thin film manifestations of the hexagonal $AB$O$_3$ oxides.  We cover the stability of the hexagonal oxides and substrates which can be used to template the hexagonal structure. We show how the thin film geometry not only allows for further tuning of the bulk-stable manganites but also the realization of metastable hexagonal oxides such as the $R$FeO$_3$ that combine ferroelectricity with weak ferromagnetic order. The thin film geometry is a promising platform to stabilize additional metastable hexagonal oxides to search for predicted high-temperature superconductivity and topological phases in this materials class. 
\end{abstract}

\maketitle

\section{\label{sec:intro}Introduction}

Complex oxides display some of the most exotic physical states known. The subtle interplay of Coulomb interactions, electron-lattice coupling and spin/orbital ordering gives rise to phenomena as diverse as high-temperature superconductivity and ferromagnetism. Synthesizing complex oxides in the thin film form offers further opportunities to tune the ground state. Here, strain imparted from a substrate, dimensionality in a superlattice architecture, or charge transfer/coupling at an interface can unleash further emergent properties not present in the parent compounds\cite{hwang2012emergent, schlom2008thin}.  Moreover, complex oxide thin films not only offer opportunities to study diverse physical phenomena but could be harnessed for a number of next-generation applications\cite{mannhart2010oxide}.

\begin{figure}
    \centering
    \includegraphics[width = \columnwidth]{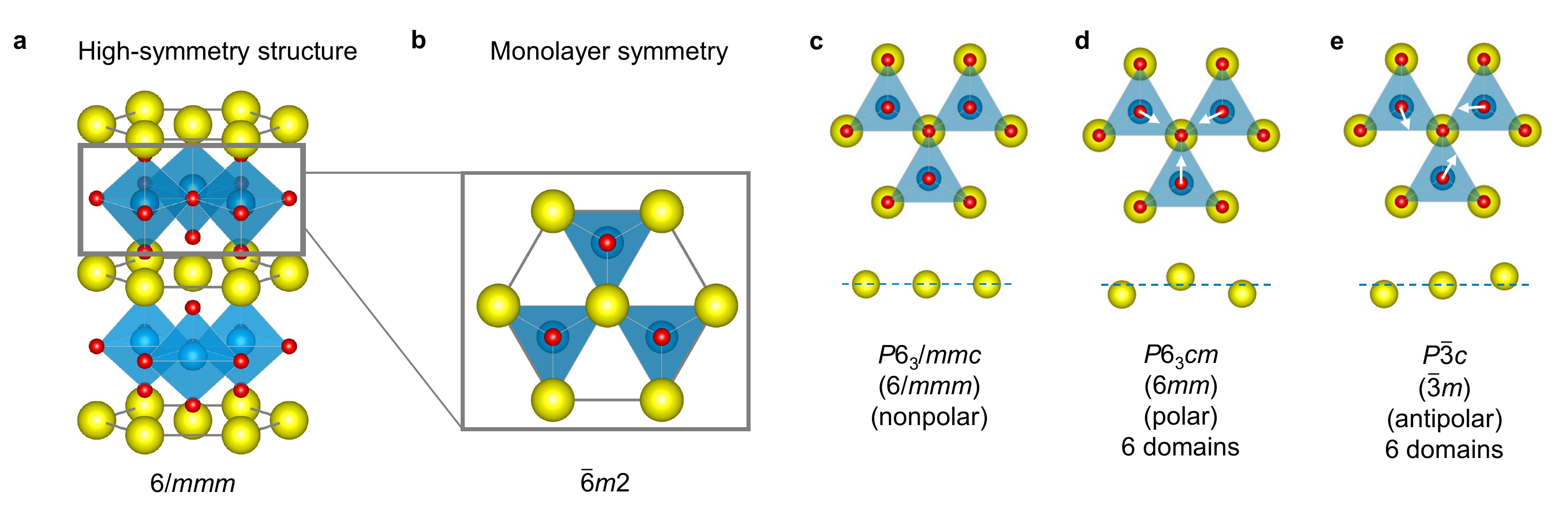}
    \caption{Crystal structure and symmetry of the hexagonal $AB$O$_3$ compounds. The $A$-site, $B$-site and oxygen atoms are shown in yellow, blue and red, respectively. (a) The parent centrosymmetric $P$6$_3$/$mmc$ structure that is non-polar with point symmetry 6/$mmm$ .  (b) A slice of the half-unit-cell plane as indicated in the box in (a), projected down the [001] crystallographic zone axis. The $B$-site cations sit on a trigonal lattice with a locally lowered (noncentrosymmetric) symmetry compared to the full unit cell structure, which consists of two of these $B$-site planes rotated 60\degree\ with respect to each other, sandwiched between the triangular rare-earth layers. (c)-(e) Displacement patterns found in hexagonal $AB$O$_3$ compounds and the corresponding unit-cell space (point) symmetry groups. (c) The nonpolar parent structure, for reference. (d) The polar phase has a coordinated tilting of three $B$O$_5$ trigonal bipyramids towards a trimerization center accompanied by a ``down-up-down" displacement pattern of A-site ions along the $c$ axis. (e) The antipolar phase consists of intermediate tilt angles 30\degree\ away from the polar structure and a "down-middle-up" displacement pattern on the A-site along the $c$ axis.}
    \label{fig:structure}
\end{figure}

To date, however, much of the work on complex oxides in the thin film form has focused on cubic perovskite oxides. This review focuses on a different class of oxides with the same $AB$O$_3$ stoichiometry ($A$, $B$ = cations), the hexagonal oxides. As shown in Fig \ref{fig:structure}(a), the crystal structure differs from that of the cubic perovskites: the $B$-site cation is surrounded by a trigonal bipyramid arrangement of oxygen atoms in contrast to the oxygen octahedra characteristic of the perovskite oxides. Planes of corner-sharing trigonal bipyramids are layered with planes of the $A$-site cation in a quasi-two-dimensional structure. (We refer the reader to an excellent recent review\cite{nguyen2020hexagonal} that covers a more generic class of hexagonal oxides that include face-sharing polyhedra and other structural types.) We focus on this class of materials not to generically expand the study of oxide compounds, but specifically because the distinct symmetry and the crystal field environment can offer unique opportunities to realize novel properties not present in the cubic perovskite oxides.

\begin{figure}
    \centering
    \includegraphics[scale = .62]{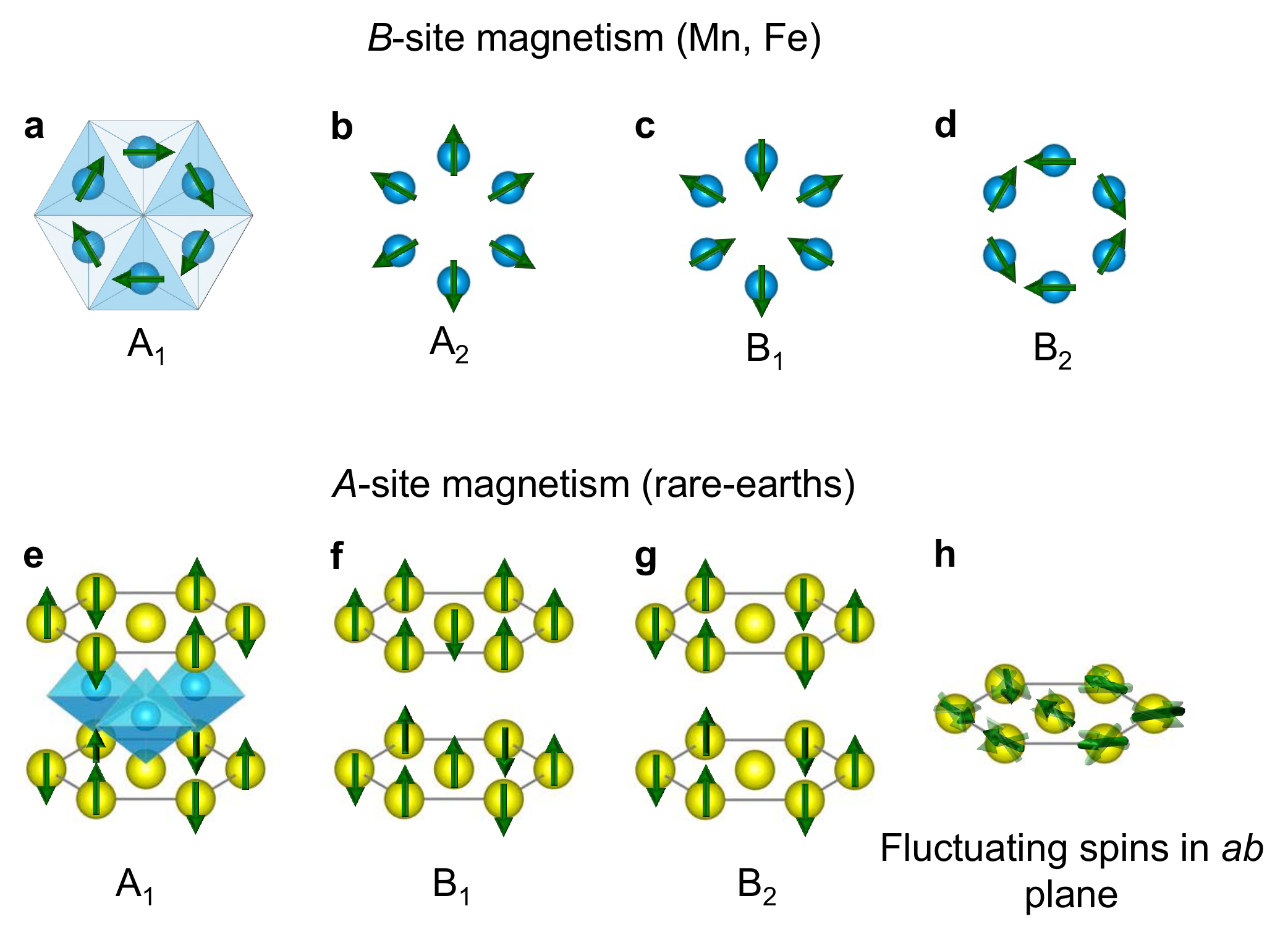}
    \caption{Magnetic order in the hexagonal $AB$O$_3$ compounds. (a)-(d) Spin orientation on the $B$-site. (e)-(h) Spin orientation on the $A$-site, for $A$ = rare-earth element. The fully ferromagnetically ordered A$_2$ configuration is not shown. (h) In the absence of $B$-site magnetism, the geometric frustration experienced by the rare-earth moments on the triangular lattice can lead to suppressed ordering temperatures or spin fluctuations persistent to the lowest temperatures.}
    \label{fig:magnetism}
\end{figure}

While considerably less studied than the perovskite oxides, the hexagonal $AB$O$_3$ materials also display a rich array of physical phenomena (see overview in Table \ref{tab:properties}). The most commonly studied hexagonal oxides are the rare earth manganites, $R$Mn$O_3$ where $R$ = Sc, Y, In, Dy–Lu. These materials display robust improper ferroelectricity well above room temperature\cite{yakel1963crystal,Lilienblum2015ferroelectricity} as a consequence of a lattice trimerization involving a coordinated tilting of the MnO$_5$ bipyramids and a distortion of the rare earth ion layers\cite{van2004origin, fennie2005ferroelectric}, see Fig. \ref{fig:structure}.  Notably, this ferroelectricity coexists and is coupled with antiferromagnetic order\cite{bertaut1963proprietes, Huang1997,Fiebig2002,lottermoser2004magnetic,Lorenz2013hexagonal}, as shown in Fig. \ref{fig:magnetism}. More recently, the hexagonal $R$FeO$_3$ compounds have been studied\cite{akbashev2011weak, wang2013room, disseler2015magnetic}: in these compounds there is a proposed coupling of ferroelectricity with weak ferromagnetic order\cite{das2014bulk}. In addition to proposed uses in energy-efficient magnetoelectric spin–orbit logic devices\cite{manipatruni2019scalable}, the ferroelectricity and ferroelectric domain walls in the hexagonal manganites have emerged as fascinating model systems to study diverse physical phenomena.  The ferroelectric domain walls display tunable metallic conductivity\cite{choi2010insulating,meier2012anisotropic,mundy2017functional}. The emergence of the ferroelectric domain structure, exhibiting a topologically protected vortex pattern, is further a platform to explore the Kibble-Zurek framework and spontaneous symmetry breaking in a condensed matter system\cite{griffin2012scaling, lin2014topological, meier2017global}. The structural phase transitions also have been proposed to display Higgs and Goldstone physics \cite{meier2020manifestation}.

\begin{figure}
    \centering
    \includegraphics[width = \columnwidth]{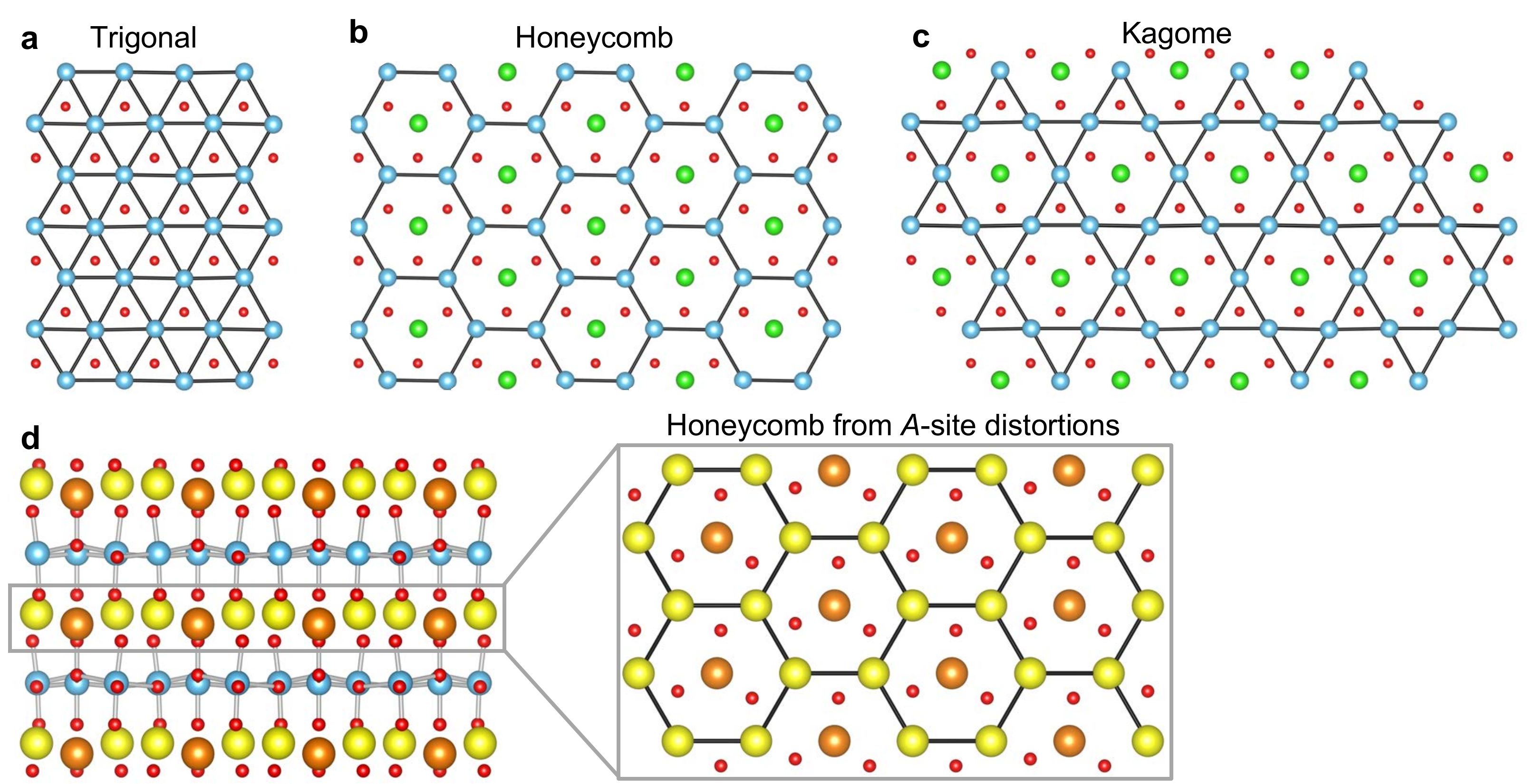}
    \caption{Symmetry embedded in the $AB$O$_3$ structure. (a) A slice of the $B$-site plane from the $AB$O$_3$ structure, projected down the [001] crystallographic zone axis.  The $B$-site cations follow a trigonal pattern. (b),(c) Replacement of the $B$-site cation with a different element at 1/4 or 1/3 filling can construct a honeycomb or kagome lattice, respectively. (d) The honeycomb lattice is also formed on the $A$-site of the P6$_3cm$ structure where the blue atoms are displaced up and the green atoms displaced down in the "up-up-down" polarization direction.  Oxygen in red, $A$-site cations in yellow/orange and $B$-site cations in blue/green.}
    \label{fig:lattice}
\end{figure}

In addition to multiferroics, the hexagonal $AB$O$_3$ family offers additional prospects to stabilize emergent magnetic ground states including the elusive quantum spin liquid state. In a quantum spin liquid, spins are highly correlated and strongly frustrated due to the crystal symmetry (e.g., triangular, honeycomb or Kagome). The resulting degeneracy between competing ground state spin configurations leads to a highly entangled state that resists macroscopic magnetic ordering to the lowest temperatures\cite{norman2016colloquium}. Importantly, the “spinon” quasiparticle excitations of this system can be itinerant Majorana fermions with a gapless dispersion of relevance to quantum computing\cite{kasahara2018majorana}. In addition to the intrinsic triangular symmetry of the hexagonal $AB$O$_3$ oxides, chemical doping or lattice distortions can construct a honeycomb and Kagome lattice with additional opportunities to realize frustrated magnetism. For example, InCu$_{2/3}$V$_{1/3}$O$_3$ has both Cu$^{2+}$ and V$^{5+}$ on the $B$-site lattice. These cations in the two:one ratio arrange such that the Cu$^{2+}$ forms a honeycomb lattice\cite{kataev2005structural, yan2012magnetic} as shown in Fig. \ref{fig:lattice}(b). Here, the honeycomb Cu$^{2+}$ atoms seem to order antiferromagnetically \cite{liu2013interlayer} rather than behave as a quantum spin liquid.  (This system could also be proximate to chiral superconductivity\cite{wu2013correlated}). In the LaCu$_{3/4}$Mo$_{1/4}$O$_3$ compound, the three:one ratio of Cu$^{2+}$ to Mo$^{6+}$ generates a Kagome arrangement of copper atoms\cite{vander1999la4cu3moo12} depicted in Fig. \ref{fig:lattice}(c). Finally, we note that the "up-up-down" ferroelectric trimerization on the $A$-site generates a honeycomb arrangement as well\cite{clark2019two} as shown in Fig. \ref{fig:lattice}(d). In TbInO$_3$, the resulting magnetic frustration on the terbium sub-lattice leads to a lack of order to the lowest temperatures and TbInO$_3$ is a promising quantum spin liquid candidate \cite{clark2019two}.

Finally, the hexagonal oxides have distinct oxygen sublattices to their perovskite counterparts. In contrast to the oxygen octahedra characteristic of the perovskites, the trigonal bipyramid intrinsic to the hexagonal $AB$O$_3$ structure has two distinct oxygen atoms located at the apical and in-plane coordinations.  These oxygen atoms can have different bonding to the transition metal atom\cite{mundy2012atomic}, which in principle can be harnessed for oxygen ion conductivity\cite{remsen2011synthesis}. There have also been theoretical predictions that the crystal field splitting surrounding the transition metal oxide could be harnessed to stabilize superconductivity in hexagonal nickelates (mimicking the electronic structure of the superconducting cuprates and pnictides)\cite{hu2015predicting}.  

\begin{center}
\begin{table}
 \begin{tabular}{||c c c||} 
 \hline
 Compound & Properties  & References \\ [0.5ex] 
 \hline\hline
$R$Mn$O_3$, $R$ = Sc, Y,  Dy–Lu  & Multiferroic & \cite{lottermoser2004magnetic} \\ 
$R$Mn$O_3$, $R$ = Sc, Y,  Dy–Lu  & Photovoltaic & \cite{huang2015hexagonal,Han2015switchable} \\ 
 YIn$_{1-x}$Mn$_x$O$_3$ & Blue pigment & \cite{li2016determination}  \\
 DyMnO$_3$ & Ion/electron conductor & \cite{remsen2011synthesis}  \\
$R$MnO$_3$ & oxygen storage & \cite{skjaervo2016interstitial, remsen2011synthesis} \\ 
 \hline
 $R$FeO$_3$, $R$ = Lu, Yb, Sc & Multiferroic & \cite{akbashev2011weak, wang2013room, disseler2015magnetic}  \\
 DyFeO$_3$ & Antiferroelectric & \cite{kasahara2021room} \\
 YFeO$_3$, InFeO$_3$ & Photocatalyst/water splitting & \cite{wu2004selective, zhang2012controllable, zhang2020infeo3}\\
 \hline
 $R_2$CuTiO$_6$, $R$ = Y, Dy, Ho, Er, and Yb & High-$\kappa$ dielectric & \cite{choudhury2010electric}  \\
  \hline
 TbInO$_3$ & Quantum spin liquid candidate & \cite{clark2019two, kim2019spin}  \\

 \hline
YCrO$_3$, YVO$_3$ & Predicted topological semi-metal & \cite{weber2019topological}  \\
\hline
YNiO$_3$ & Predicted superconductor & \cite{hu2015predicting, lu2018d+}  \\

 \hline
\end{tabular}
\caption{Observed and predicted properties of the $AB$O$_3$ hexagonal materials.}
\label{tab:properties}
\end{table}
\end{center}

Thin film manifestations of these hexagonal oxides offer further opportunities to not only scale down the materials to fundamental thickness limits\cite{nordlander2019ultrathin} and heterostructure them to realize additional functional properties\cite{mundy2016atomically}, but furthermore to exploit epitaxy and layering\cite{garten2021stromataxic} to stabilize metastable compounds. Thin film epitaxy of the hexagonal $AB$O$_3$ oxides offers additional challenges in comparison to the cubic perovskite oxides.  In this review, we first summarize the stability of the hexagonal $AB$O$_3$ compounds.  We then describe the thin film deposition of the bulk-stable $R$MnO$_3$ compounds, including the lattice matching of the hexagonal crystal structure to commercially available substrates. We finally discuss the use of epitaxy to expand the stability of this phase. 

\section{\label{sec:structure}Stability of the hexagonal ABO$_3$ phase}

Compounds with the $AB$O$_3$ stoichiometry can form a variety of cubic, orthorhombic, and hexagonal phases. The formation of the hexagonal phase is dictated by both structural stability and the electronic energy of the $B$-site in the trigonal bipyramid oxygen coordination complex.  The structural stability of the competing cubic perovskite phase can be estimated with the tolerance factor ($t$), a geometric quantity based on the ionic radii ($r$) of constituent atoms that indicates how well a given $AB$O$_3$ compound fits in the cubic perovskite structure. Here, $t$ is given by: 
$$
t = \frac{(1/\sqrt{2}) \text{face diagonal}}{\text{unit cell length}} =  \frac{r_A+r_O}{\sqrt{2}(r_B+r_O)}
$$

\begin{figure}
    \centering
    \includegraphics[width = \columnwidth]{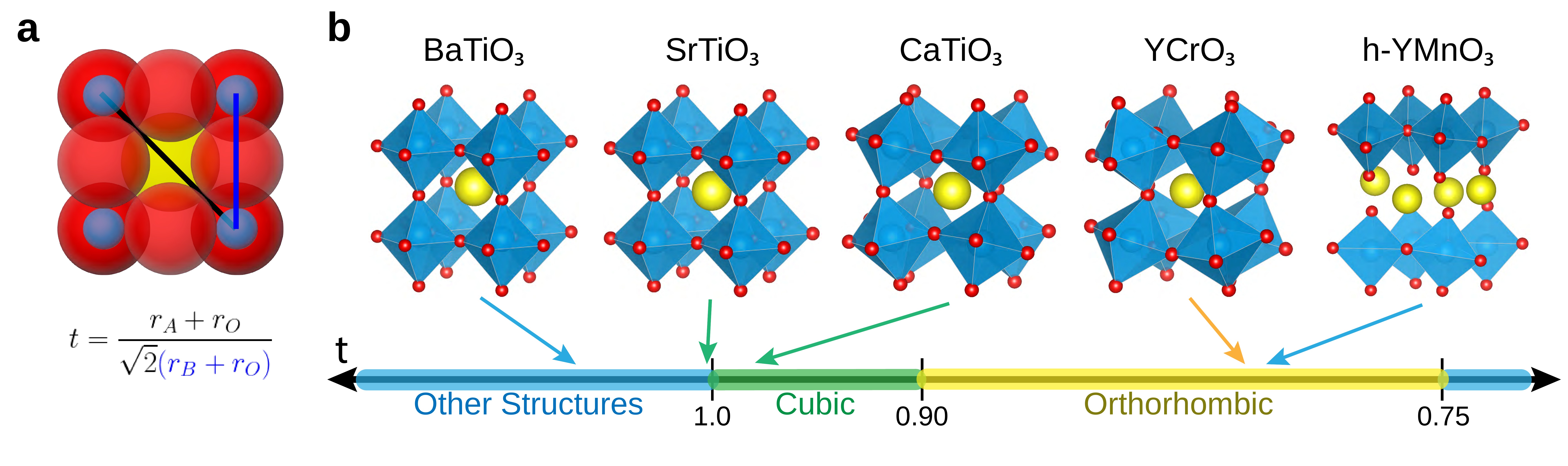}
    \caption{The perovskite Goldschmidt tolerance factor. (a) The tolerance factor, $t$, is calculated as the ratio of $\frac{1}{\sqrt{2}}$ times a face diagonal (2r$_A$ + 2r$_O$) and a unit cell length (2r$_B$ + 2r$_O$). $t=1$ for ions which perfectly pack into the cubic perovskite structure. (b) Examples of $AB$O$_3$ compounds with various tolerance factors and the approximate regions of stability for each structure. As $t$ decreases from 1, the $B$-site octahedral cages tilt and bond lengths change forming successively lower-symmetry structures. The cubic perovskite phase with space group $Pm\overline{3}m$ occurs for approximately $0.90<t<1.0$, while the orthorhombic phase with space group $Pnma$ or $Pbnm$ occurs for $0.75<t<0.90$ \cite{hines1997atomistic}. Outside of this range, other structures including a hexagonal phase with space group $P6_3cm$, a tetrahedral phase, and the hexagonal ilmenite structure are formed. The tolerance factors for BaTiO$_3$, SrTiO$_3$, CaTiO$_3$, YCrO$_3$, and YMnO$_3$ are 1.07, 1.01, 0.97, 0.83, and 0.81, respectively.}
    \label{fig:tolerancefactor}
\end{figure}

The tolerance factor is 1 for $A$ and $B$ ions which can be packed into a perfect cubic perovskite. The well-known perovskite SrTiO$_3$ has $t$=1.01, when calculated with r$_A$ = r$_{Sr^{2+}(XII)}$ = 144 pm, r$_B$ = r$_{Ti^{4+}(VI)}$ = 60.5 pm, and r$_O$ = r$_{O^{2-}(II)}$ = 135 pm where r$_{Sr^{2+}(XII)}$ is the ionic radius of Sr$^{2+}$ with twelve-fold oxygen coordination \cite{shannon1970revised}.   As $t$ deviates from 1, the structure is distorted from a cubic perovskite (Fig. \ref{fig:tolerancefactor}). For example, YCrO$_3$ with $t$=0.83, forms a distorted orthorhombic perovskite phase. A tolerance factor far from 1 does not imply the stability of the hexagonal phase, but instead identifies candidate compounds that are less stable in the cubic perovskite phase. YMnO$_3$ and YFeO$_3$, both with $t$=0.81, have less structural stability in the cubic perovskite phase. 

\begin{figure}
    \centering
    \includegraphics[scale=1]{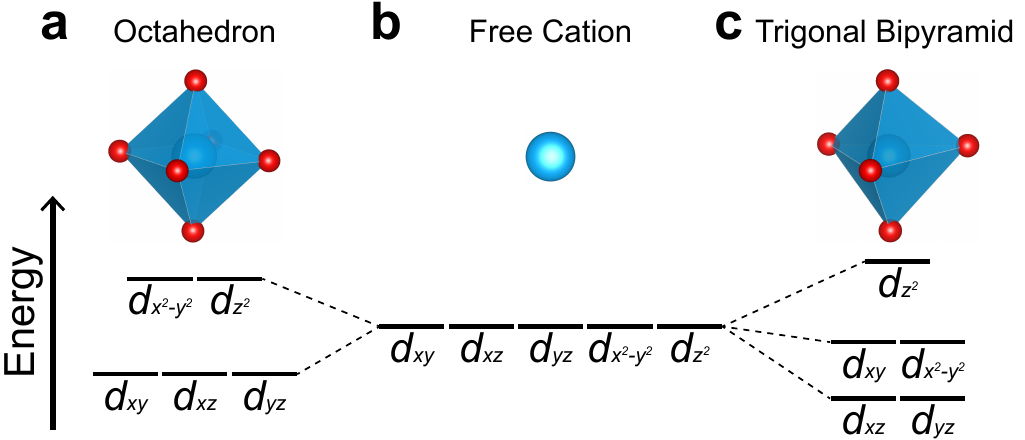}
    \caption{The $d$-orbital energy configuration for ions with (a) octahedral oxygen coordination (as for $B$-site cations in cubic perovskites), (b) free ions, and (c) trigonal bipyramid coordination (as for $B$-site cations in hexagonal ABO$_3$ materials).}
    \label{fig:CEFsplitting}
\end{figure}

\begin{figure}
    \centering
    \includegraphics[scale = 0.6]{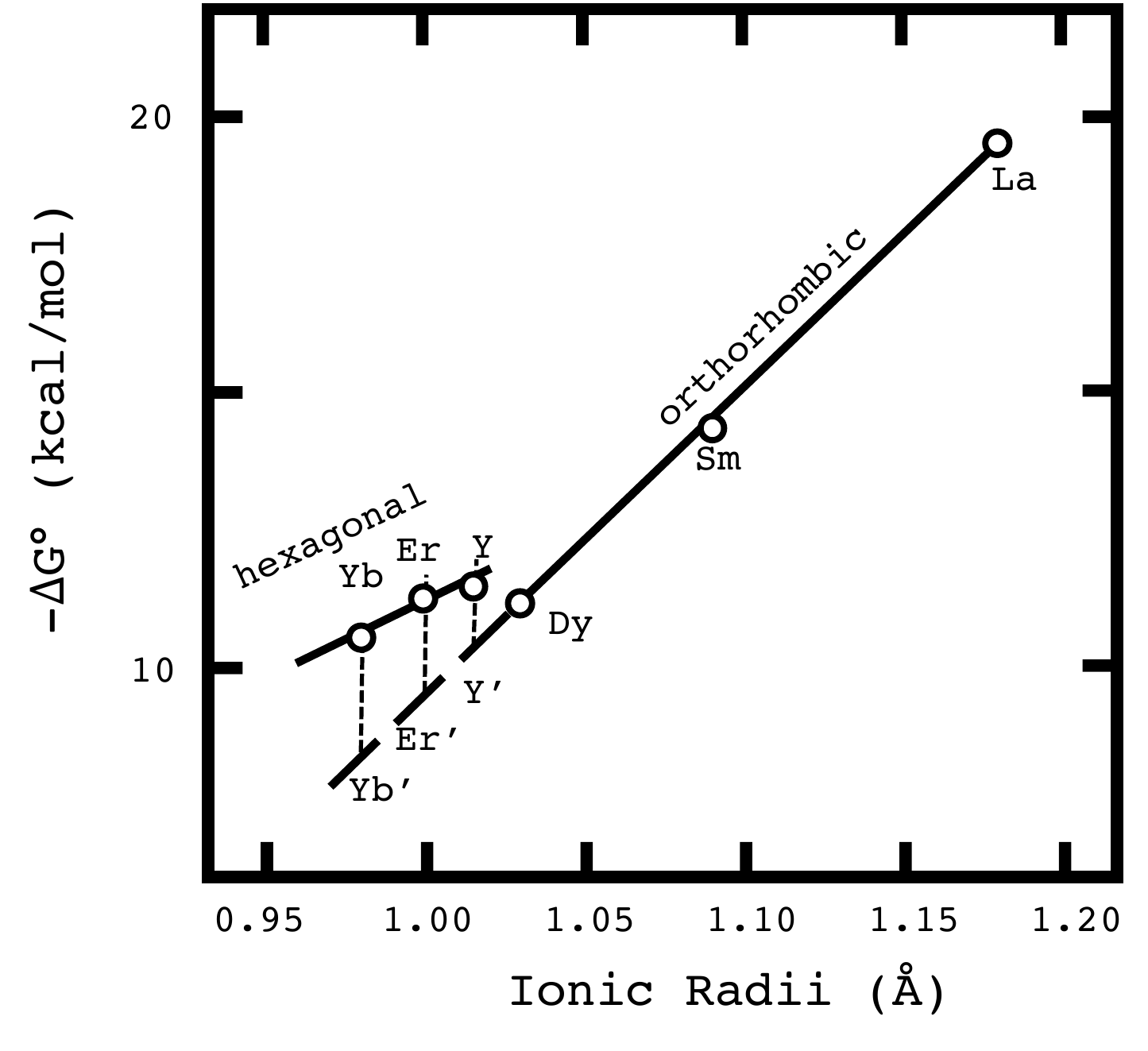}
    \caption{The dependence of free energy of formation on rare-earth ionic radii, RE$^{3+}$ in $R$MnO$_3$, at 1200$^\circ$. RE' indicates the estimated free energy of formation for the perovskite structure in high pressure. Reprinted with permission from Kamata \textit{et al}. Mater. Res. Bull. \textbf{14}, 1007-1012 (1979). Copyright 1979 Elsevier.\cite{kamata1979thermogravimetric}}
    \label{fig:freeenergy}
\end{figure}

Despite having the same tolerance factor, YFeO$_3$ assumes an orthorhombic perovskite structure whereas YMnO$_3$ crystallizes in a hexagonal phase.  In addition to the structural stability of competing phases, electronic stability impacts the bulk stable crystal structure. The electronic stability of the hexagonal oxides is based on the crystal field splitting of the $d$-orbitals of the $B$-site transition metal ions with trigonal bipyramid (five-fold) oxygen coordination. With this ligand geometry, the five $d$-orbitals split into three energy levels: two doubly degenerate lower energy levels and one non-degenerate high energy level (Fig. \ref{fig:CEFsplitting}). In contrast, a perovskite with octahedral (six-fold) coordination on the $B$-site has two energy levels: the three-fold degenerate t$_{2g}$ low energy level and the doubly degenerate e$_g$ high energy level. Fe$^{3+}$, with the electron configuration [Ar]3d$^5$, is more stable in the octahedral coordination, which avoids occupying the highest energy level in the trigonal bipyramid configuration. As a result, $R$FeO$_3$ compounds are more stable as orthorhombic perovskites. In contrast, Mn$^{3+}$, with configuration [Ar]3d$^4$, is more stable in a trigonal bipyramid complex where the four valence electrons populate the four lower energy orbitals. With octahedral coordination, Mn$^{3+}$ forms a degenerate high-spin state with its fourth valence electron in either of the two e$_g$ orbitals; these perovskite manganites are Jahn-Teller active \cite{reinen1979jahn}. The electron in the higher energy orbitals lowers the stability of the compound. Thus, the $R$MnO$_3$ compounds form a hexagonal crystal structure for small $R$ ($R$ = Sc, Y, Dy-Lu) as shown in Fig. \ref{fig:freeenergy}. 

\begin{figure}
    \centering    \includegraphics[width = \columnwidth]{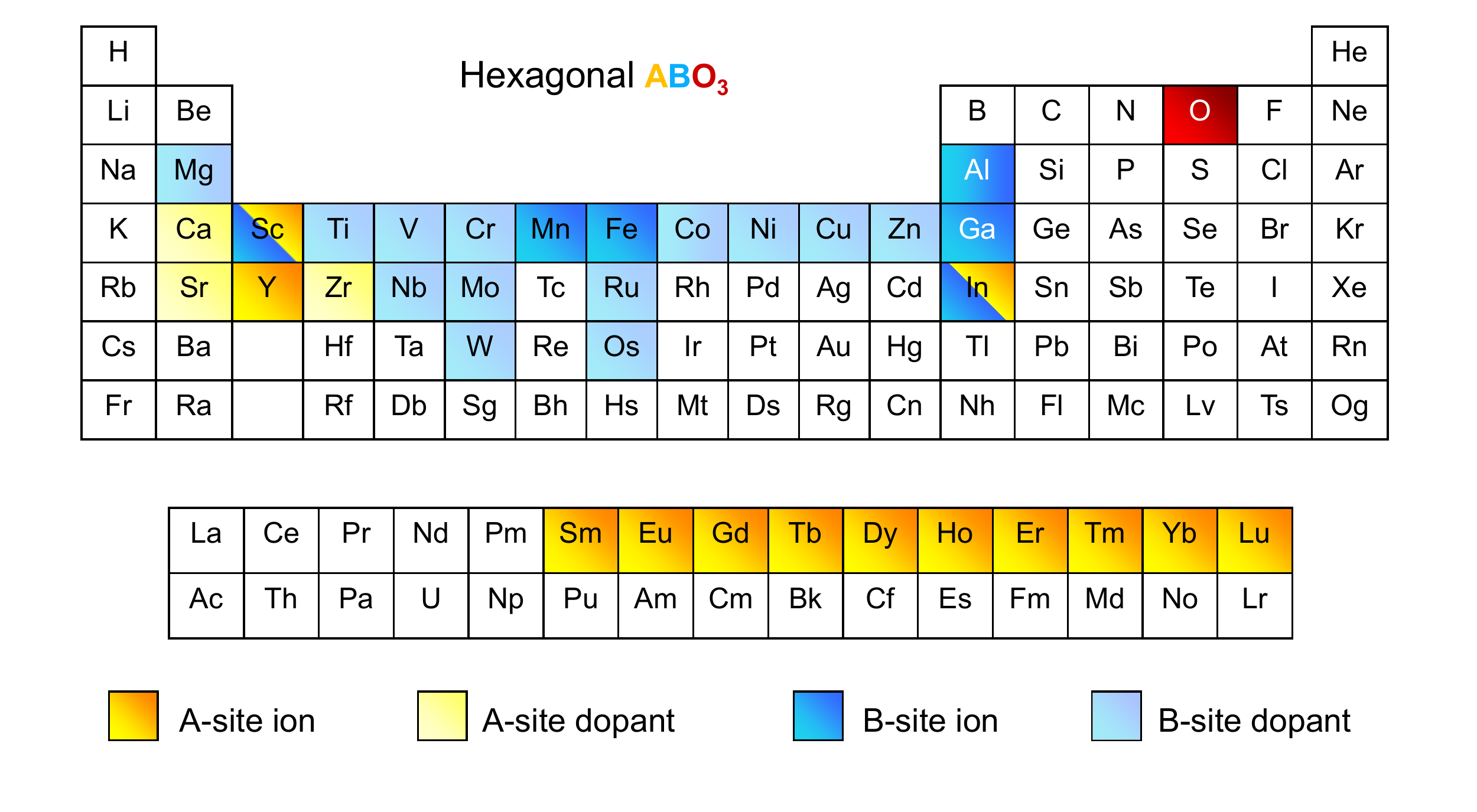}
    \caption{Periodic table indicating the elements that can be stabilized in the hexagonal $AB$O$_3$ phase. $A$-site ions are colored in yellow and $B$-site ions are colored in blue.  We color in light blue/yellow the elements which can be found as dopants or to partially occupy the site.}
    \label{fig:periodic}
\end{figure}

In Fig. \ref{fig:periodic}, we summarize the elements which have been found in the hexagonal $AB$O$_3$ polymorph.  In contrast to the cubic perovskites where almost every element on the periodic table can occupy one of the three lattice sites\cite{ hellwege1979landolt, schlom2008thin}, the hexagonal structure can form with more limited chemical compositions. Only six cations are known to fully occupy the $B$-site although there are much wider range of elements which can be stabilized as dopants or as partial occupants of this lattice site.  We note that thin film stabilization is a powerful platform for synthesizing phases which are metastable as bulk crystals\cite{nagashio2002metastable}. While YMn$_{1-x}$Fe$_x$O$_3$ could be stabilized in the P6$_3cm$ structure for $x<$0.3 in bulk crystals, epitaxial stabilization has led to the construction of LuFeO$_3$ with the hexagonal P6$_3cm$ structure\cite{akbashev2011weak, wang2013room, disseler2015magnetic, katayama2017epitaxial}.  Explicit stromataxy -- precise control over the layering -- has enabled further stability of this phase\cite{garten2021stromataxic}.  

\section{\label{sec:substrates}Substrate templates for hexagonal ABO$_3$ thin films}
\begin{figure}
    \centering
    \includegraphics[width = \columnwidth]{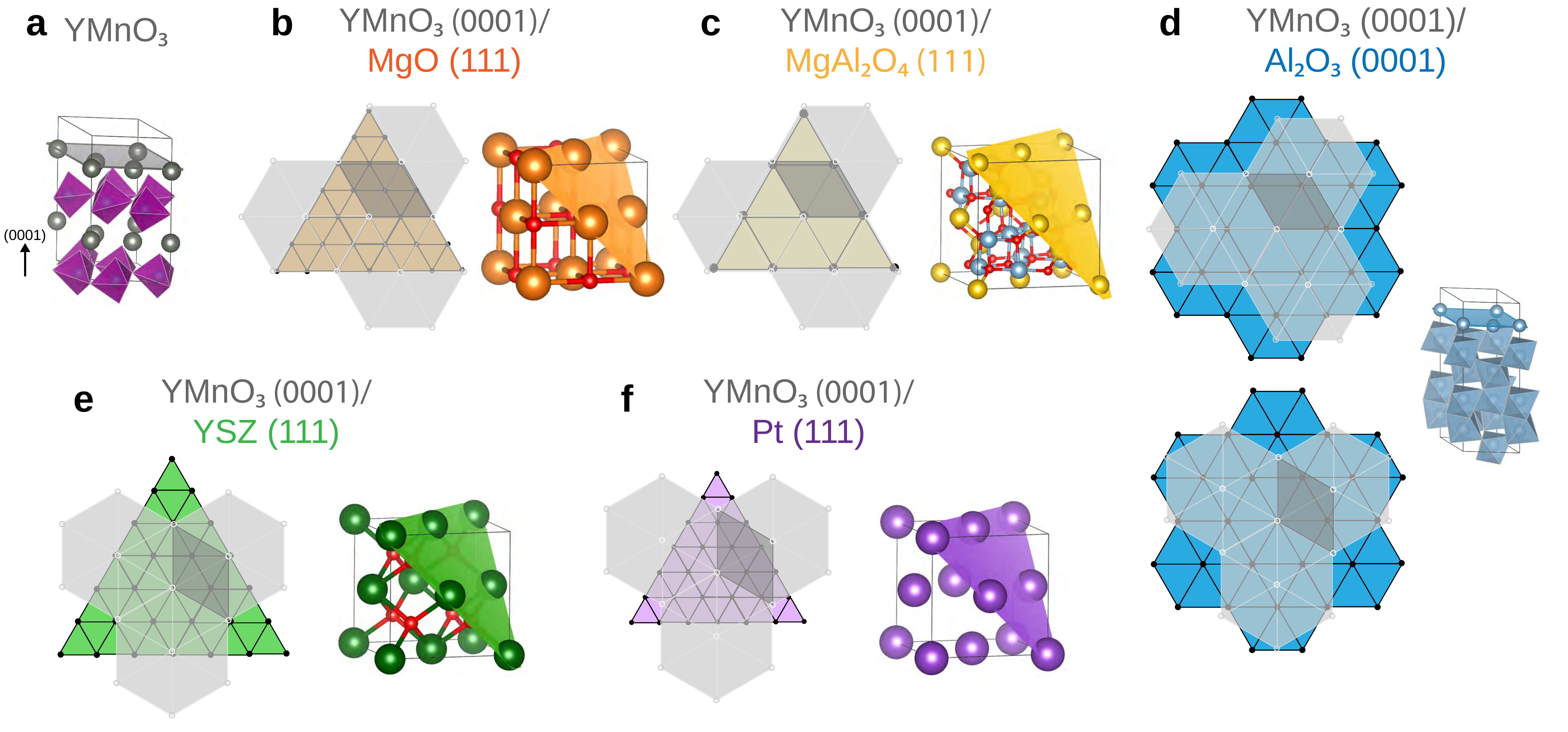}
    \caption{Lattice matching between YMnO$_3$ and common substrates. (a) The structure of YMnO$_3$ with a (0001) plane unit cell of yttrium atoms highlighted in grey. (b) Lattice matching between yttrium atoms in the (0001) plane of YMnO$_3$ in grey and the magnesium atoms in the (111) plane of MgO in orange. Lattice mismatch (with a factor of 2): -2.9\%. (c) Lattice mismatch between YMnO$_3$ in grey and the magnesium atoms of MgAl$_2$O$_4$ (111) in yellow. Lattice mismatch: -7.0\%. (d) Lattice mismatch of YMnO$_3$ with the blue aluminum atoms of Al$_2$O$_3$(0001) in two orientations: aligned (Lattice mismatch (with a factor of $\frac{4}{3}$): 3.4\%) and with a 30$^\circ$ rotation (Lattice mismatch (with a factor of $\frac{2}{3}$): -10.5\%). (e) Lattice mismatch between 30$^\circ$ rotated YMnO$_3$ and the zirconium atoms of yttria-stabilized zirconia (YSZ) in green. Yttrium is omitted from the YSZ crystal structure for clarity. Lattice mismatch: 2.6\%. (f) Lattice mismatch between 30$^\circ$ rotated YMnO$_3$ and the platinum atoms of Pt(111). Lattice mismatch (with a factor of $\frac{4}{3}$): 4.6\%.}
    \label{fig:latticematching}
\end{figure}

\begin{figure}
    \centering
    \includegraphics[width=\columnwidth]{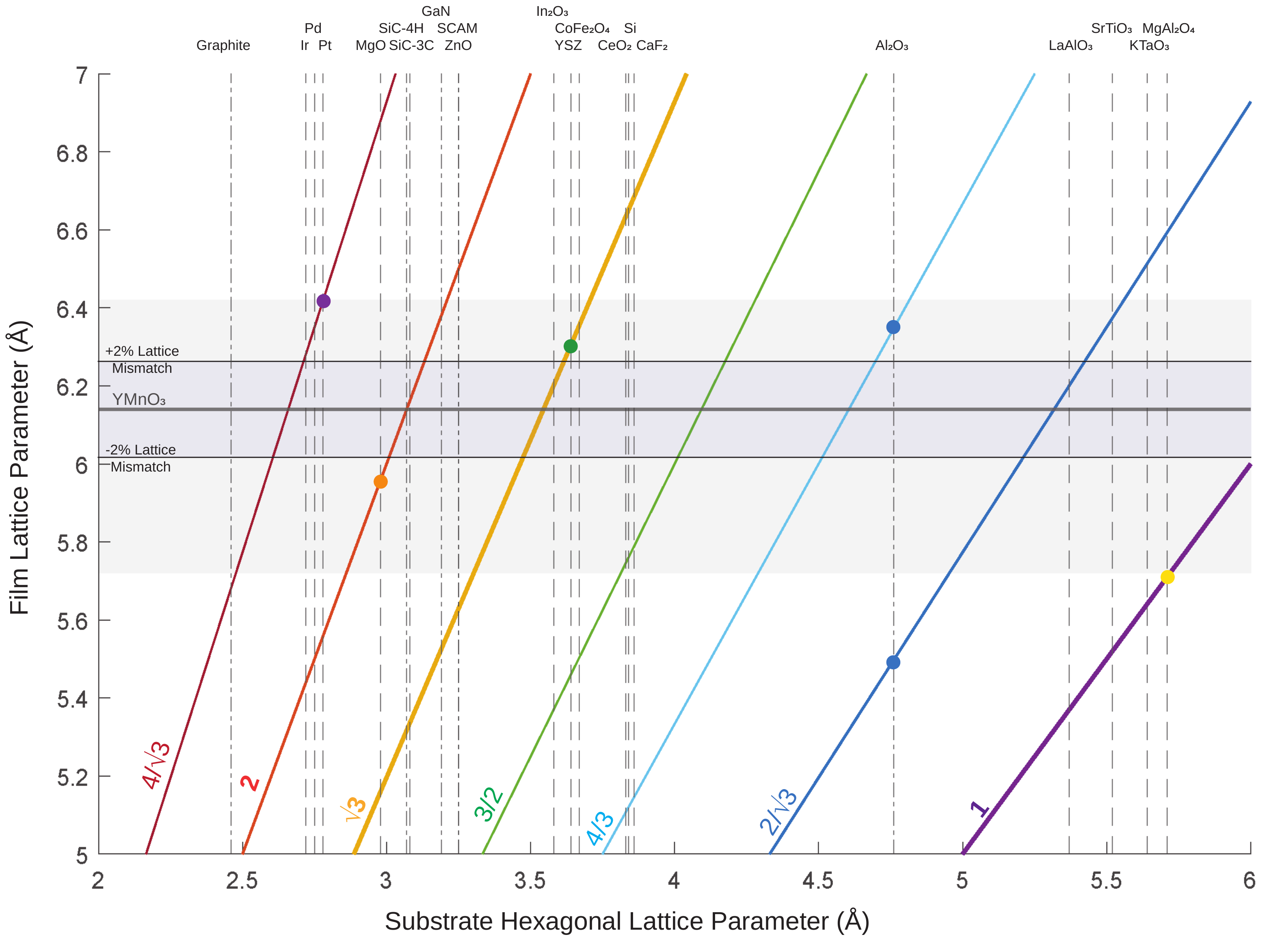}
    \caption{In-plane lattice matching for hexagonal materials. Vertical lines indicate substrate (0001) (dot-dashed lines) and (111) (dashed lines) in-plane lattice parameters. The intersection between a vertical line and scaling factor indicates a possible effective substrate lattice parameter. The horizontal line at 6.14 \AA \ represents YMnO$_3$ and the boundaries for $\pm 2\%$ lattice mismatch are labelled. The grey region indicates typical hexagonal film lattice parameters. Colored dots correspond to the lattice matching diagrams in Fig. \ref{fig:latticematching}. The thickness of the scaling factor lines roughly matches their likelihood of being realized, with 1 and $\sqrt{3}$ the most common.}
    \label{fig:numberline2d}
\end{figure}

While sputtering, metal-organic chemical vapor deposition (MOCVD), molecular-beam epitaxy (MBE) and pulsed laser deposition (PLD) have emerged as powerful tools to synthesize oxide materials in the film form, there are unique challenges to the deposition of hexagonal oxides.  In contrast to the more commonly studied perovskite oxides, where there is a ``menu" of isostructural, commercially available perovskite substrates\cite{uecker2017large} with various lattice constants, the most readily available oxide substrate used to stabilize hexagonal films is Al$_2$O$_3$. This substrate is not well lattice-matched to all desired hexagonal $AB$O$_3$ films.  In addition to hexagonal oxides, cubic substrates can be used in the (111) orientation. Figure \ref{fig:latticematching} shows the mismatch between the prototypical YMnO$_3$ and (0001) Al$_2$O$_3$, (111) cubic oxides MgO, MgAl$_2$O$_4$, (ZrO$_2$)$_{0.905}$(Y$_2$O$_3$)$_{0.095}$ (9.5.mol\% yttria-stabilized zirconia, YSZ), and (111) oriented metallic platinum.  In addition to directly aligning on the substrate, the YMnO$_3$ could also adopt a 30$^\circ$ rotation with respect to the substrate orientation (this would be analogous to the 45$^\circ$ rotation cubic perovskites might assume in the (001) direction). Figure \ref{fig:numberline2d} summarizes the lattice matching between the P6$_3cm$ YMnO$_3$ and many commerically available substrates.  While (111)-oriented cubic perovskites also have the correct symmetry, our experience is that these substrates seed the (111)-perovskite film rather than the intended hexagonal polymorph. 

\section{\label{sec:bulkstable} Epitaxy of bulk-stable compounds}

\begin{figure}
    \centering
    \includegraphics[scale =0.45]{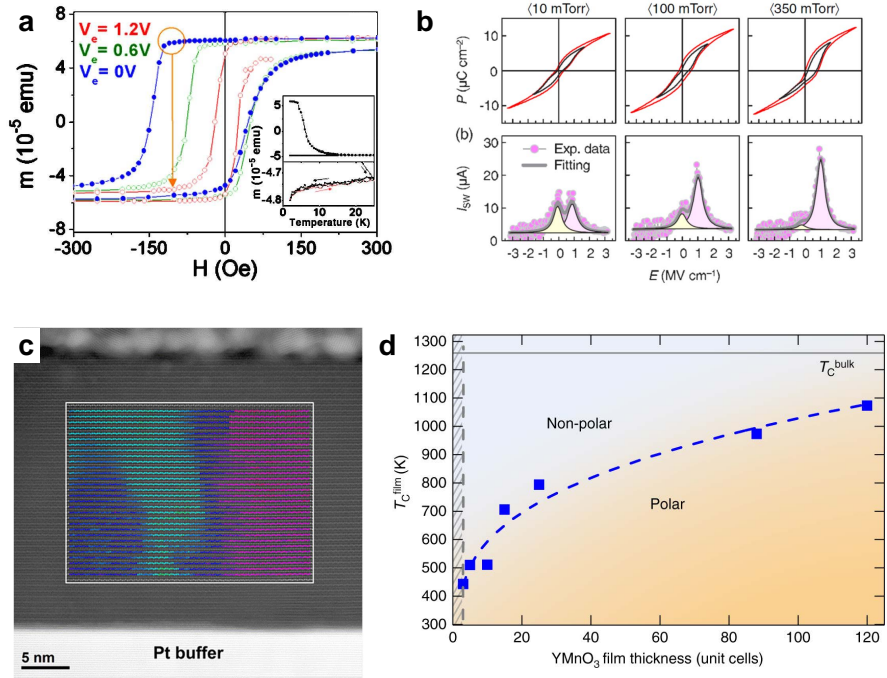}
    \caption{Ferroelectric properties of hexagonal $R$MnO$_3$ thin films. (a) Voltage control of exchange bias in a permalloy layer through interfacial magnetic exchange in epitaxial multiferroic YMnO$_3$ grown on Pt(111)/SrTiO$_3$(111). Reprinted figure with permission from Laukin \textit{et al}. Phys. Rev. Lett. \textbf{97}, 227201 (2006). Copyright (2006) by the American Physical Society.\cite{Laukhin2006electric} (b) Modification of ferroelectric hysteresis loops in hexagonal HoMnO$_3$ films on Pt(111)/Al$_2$O$_3$(0001) caused by the strain-gradient-induced flexoelectric effect. Reprinted figure with permission from Lee, D \textit{et al}. Phys. Rev. Lett. \textbf{107}, 057602 (2011). Copyright (2011) by the American Physical Society.\cite{Lee2011giant} (c) Improper ferroelectric domain pattern mapped at the atomic scale in hexagonal YMnO$_3$ on Pt(111)/YSZ(111) using HAADF-STEM. Each color represents one of six total trimerization domain states, with polarization pointing either up or down. (d) Improper ferroelectric transition temperatures in hexagonal YMnO$_3$ films grown on YSZ(111) as function of film thickness. (c) and (d) reprinted from Nordlander \textit{et al}. Nat. Commun. \textbf{10}, 5591 (2019). Copyright 2019 The Authors, licensed under a  \href{http://creativecommons.org/licenses/by/4.0}{CC BY 4.0} license.\cite{nordlander2019ultrathin}}
    \label{fig:RMOpolarization}
\end{figure}

Epitaxial thin-film synthesis of hexagonal $AB$O$_3$ materials was first achieved for the prototypical family of hexagonal rare-earth manganites ($R$MnO$_3$). In particular, initial focus centered on epitaxy of YMnO$_3$ as a model system for this materials class and YMnO$_3$ remains the most intensely studied member of hexagonal $AB$O$_3$ materials. The first attempts at epitaxial synthesis of hexagonal YMnO$_3$ in 1996 were motivated by its high-temperature uniaxial ferroelectricity combined with its suitability for integration with silicon as non-volatile ferroelectric memories \cite{Fujimura1996}. Following the demonstration of magnetoelectric coupling effects in hexagonal $R$MnO$_3$ bulk crystals \cite{Fiebig2002,lottermoser2004magnetic}, interest shifted towards thin-film manifestation of such multiferroic properties and potential magnetoelectric effects. In 2006, Laukhin et al. demonstrated electrical control of magnetism in a permalloy/YMnO$_3$ thin-film heterostructure \cite{Laukhin2006electric}, see Fig.\,\ref{fig:RMOpolarization}(a). A giant flexoelectric effect was further observed in strained HoMnO$_3$ films \cite{Lee2011giant}, which could be used to tune the ferroelectric properties (Fig. \ref{fig:RMOpolarization}(b)). The discovery of the topologically protected vortex domain patterns in bulk crystal $R$MnO$_3$ compounds in 2010 \cite{choi2010insulating,Jungk2010Electrostatic} and subsequently their related domain-wall functionalities \cite{meier2012anisotropic,mundy2017functional}, sparked a renewed interest in the multiferroic domain structure and thickness scaling in hexagonal $R$MnO$_3$ thin films. Furthermore, the improper nature of the ferroelectric order in hexagonal $AB$O$_3$ oxides, which is driven by a non-ferroelectric structural distortion, suggests novel avenues for stabilizing and controlling spontaneous ferroelectric polarization in the ultrathin limit \cite{Sai2009absence,nordlander2019ultrathin}.

To date, the hexagonal phase has been realized as thin films for the entire bulk-stable rare-earth manganite series ($R$ = Dy-Lu, \cite{Fujimura1996,Dho2004epitaxial,bosak2002epitaxial,Kim2007growth,Gelard2008,Jang2008oxygen,Posadas2006epitaxial,Takahashi2006growth,Han2015switchable}) where epitaxial stabilization additionally extended the series to lighter rare-earth ions down to samarium (Sm-Tb, \cite{bosak2002epitaxial,Balasubramaniam2007,lee2006epitaxial,Mandal2020strain}) that normally would crystallize in the orthorhombic $Pnma$ phase as further discussed in Sec.\,\ref{sec:metastable}. Here, hexagonal SmMnO$_3$ has been achieved only on an isostructural YMnO$_3$ substrate \cite{Balasubramaniam2007}, whereas the other members can be grown epitaxially on commercially available substrates such as YSZ(111). In addition to the rare-earth series and yttrium, the $A$-site of hexagonal manganites can accommodate both scandium and indium. However, to our knowledge, only the (epitaxially stabilized) orthorhombic phase of ScMnO$_3$ has been reported in thin films; growth of hexagonal InMnO$_3$ remains limited to polycrystalline films reported in the literature \cite{Serrao2006imo}. In bulk InMnO$_3$, carefully tuning the defect chemistry or the thermal history of the crystals allows realizing trimerized domains of either $P$6$_{\mathrm{3}}cm$ symmetry (see Fig. \ref{fig:structure}(d)), just as in the other rare-earth manganites, or domains of the anti-polar $P\overline{\mathrm{3}}c$1 symmetry (Fig. \ref{fig:structure}(e)) seen at the domain walls between the polar trimerization domains in $R$MnO$_3$, all the while retaining the vortex domain pattern \cite{Kumagai2012,Huang2014,Griffin2017}. Epitaxial realization of single-crystalline hexagonal InMnO$_3$ thus provides an interesting opportunity for studying the thin-film manifestation of these complementary symmetry properties.

As mentioned in Sec. \ref{sec:substrates}, a major challenge in thin-film growth of the hexagonal manganites is the lack of isostructural or lattice-matching substrates. Added to this is a tendency to form crystallographic domains and defects due to close lattice matching between multiple crystallographic orientations of the hexagonal $R$MnO$_3$ structure and with the corresponding binary oxide $R_2$O$_3$ \cite{Jehanathan2010structure}. Indeed, the higher symmetry of cubic substrates such as YSZ, compared to that of the layered hexagonal $AB$O$_3$ phase, impedes nucleation of a completely single-domain crystalline film and can cause defects such as antiphase boundaries. Thus, careful attention to substrate surface termination may play an important role in reducing the occurrence of these types of defects \cite{Nordlander2020epitaxial}. The epitaxial quality of $R$MnO$_3$ films is also strongly dependent on substrate temperature during deposition. The hexagonal phase crystallizes down to 690\degree C \cite{Choi2004Bi}, however the highest crystalline quality is achieved in the range 750\degree C-900\degree C. Although hexagonal manganite thin films have been grown on a range of substrates including YSZ(111), Si(111), Pt(111), MgO(111), GaN(0001), ZnO(0001) and $c$-cut Al$_2$O$_3$ \cite{Dho2004epitaxial,Fujimura1996,Imada1998epitaxial,Imada2001ferroelectricity,Marti2007epitaxial,Laukhin2006electric,Wu2011marked,Chye2006molecular,Posadas2006epitaxial,Cheng2018interface}, resulting in various degrees of crystallinity, not nearly as many options are commercially available as for their perovskite $AB$O$_3$ counterparts.

Over the past decade, significant improvement of the thin-film crystalline quality of $R$MnO$_3$ has been achieved and epitaxial layer-by-layer thin-film growth with sub-unit-cell thickness precision has recently been demonstrated \cite{Nordlander2020epitaxial}. Such improvement of structural quality has been crucial to the investigation into the ultrathin manifestation of improper ferroelectric properties and domain structure \cite{nordlander2019ultrathin,Mandal2020strain,Pang2016preparation,Kim2016domain} in this class of materials (Fig. \ref{fig:RMOpolarization}(c)). In particular, the structural distortion transforming the non-polar $P$6$_{\mathrm{3}}/mmc$ phase to the polar $P$6$_{\mathrm{3}}cm$ phase, and leading to the secondary ferroelectric polarization, was shown to be significantly modified by substrate-interface proximity, as demonstrated by a combination of in-situ high-angle annular dark-field scanning transmission electron microscopy (HAADF-STEM) and optical second harmonic generation (SHG) \cite{nordlander2019ultrathin}. The resulting threshold thickness for room-temperature polarization in YMnO$_3$ on an insulating substrate was determined to be two unit cells (Fig. \ref{fig:RMOpolarization}(d)).

In hexagonal $R$MnO$_3$, the improper ferroelectric order coexists with antiferromagnetic order on the Mn$^{3+}$ sublattice below 70-120\,K, leading to multiferroicity. Because of the fully compensated nature of this antiferromagnetic order (i.e. lack of net magnetic moment), its thin-film manifestation has been challenging to study \cite{Fontcuberta2015multiferroic}. Although neutron diffraction on thicker films of YMnO$_3$ (thickness exceeding 400\,nm) shows N\'{e}el temperatures that closely match the bulk values \cite{Gelard2008}, similar neutron measurements on thinner films are precluded due to the limited thin-film volume. Moreover, SQUID magnetometry has revealed spin-glass states in oxygen-deficient YMnO$_3$ films, indicating the strong influence of oxygen off-stoichiometry on the magnetic order \cite{Jang2008oxygen,Cheng2016manipulation}. Hence, further work is needed to fully characterize the intrinsic magnetic state of ultrathin hexagonal $R$MnO$_3$. Additionally, it is expected that the antiferromagnetic domain size is significantly reduced in thin films compared with bulk crystals, hampering real-space characterization \cite{Kordel2009}. Thus, the enigmatic multiferroic coupling of ferroelectric and antiferromagnetic domain patterns observed in $R$MnO$_3$ bulk crystals \cite{Fiebig2002, Giraldo2021} remains a topic for future investigations in $R$MnO$_3$ thin films.

\begin{figure}
    \centering
    \includegraphics[width=\columnwidth]{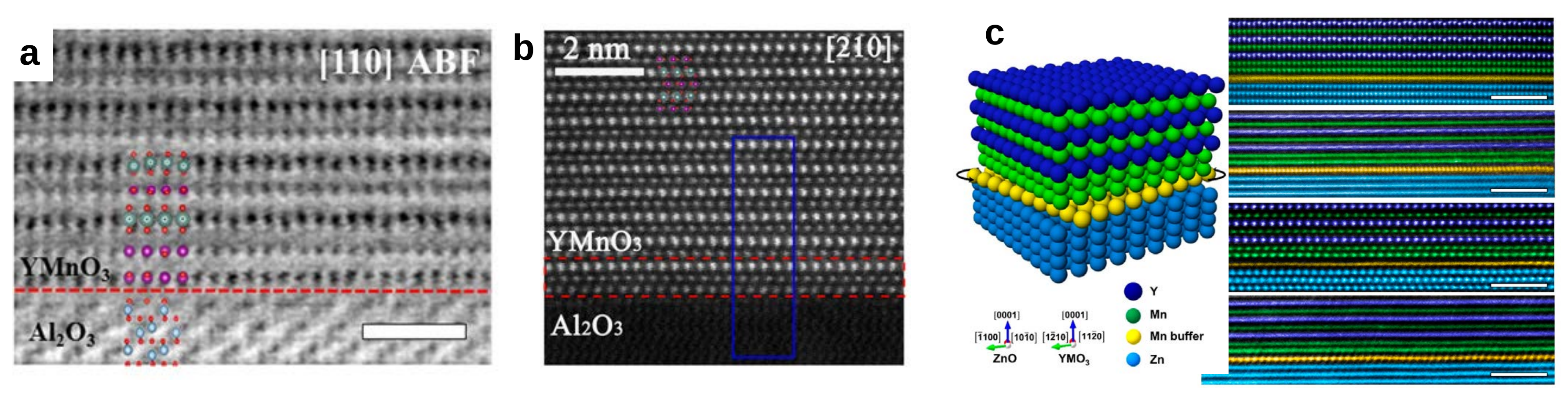}
    \caption{Substrate--film interface reconstruction in hexagonal YMnO$_3$ thin films visualized by HAADF-STEM. (a), (b) Hexagonal YMnO$_3$ films on sapphire substrates exhibit a double Mn-O layer at the interface that can host a charge ordered state of Mn$^{3+}$ and Mn$^{2+}$ ions. Reprinted from Cheng \textit{et al}. Sci. Adv. \textbf{4}, eaar4298 (2018). Copyright 2018 The Authors, some rights reserved; exclusive licensee AAAS. Distributed under a \href{https://creativecommons.org/licenses/by-nc/4.0}{CC BY-NC 4.0} license.\cite{Cheng2018interface} (c) A triple Mn-O layer is formed between hexagonal YMnO$_3$ and a Zn-terminated ZnO(0001) substrate. The Mn-O layer closest to the substrate is reported to adopt a 5\degree\ in-plane rotation relative to both film and substrate, influencing the strain relaxation of the YMnO$_3$ film. Reprinted with permission from Zhang \textit{et al}. Nano Lett. \textbf{21}, 6867-6874 (2021). Copyright (2021) American Chemical Society.\cite{Zhang2021controlling}}
    \label{fig:RMOinterface}
\end{figure}

In addition to dimensionality scaling and heterostructure integration, thin-film realization of quantum oxide materials offers the opportunity to use epitaxial constraints to further tune their functionality. The lattice mismatch with respect to the substrate can impart epitaxial strain in the thin film, or induce interface defects, that affect both electronic and magnetic properties. In hexagonal $R$MnO$_3$, strain engineering can in principle be used to tune the improper ferroelectricity in terms of both domain configuration and polarization magnitude  \cite{Artyukhin2014,Wang2014,Tan2016}. Although thin-film $R$MnO$_3$ is often grown on substrates with large lattice mismatch exceeding 2\% (see Fig. \ref{fig:numberline2d}), the resulting epitaxial strain is not obvious \cite{Dubourdieu2007thin}. Rather than inducing a coherently strained thin-film lattice, several other mechanisms are often at play that accommodate this mismatch. For example, misfit dislocations are frequently seen at the substrate--film interface, allowing bulk-like lattice constants to persist in the thin-film limit \cite{nordlander2019ultrathin}. Structural or chemical mismatch at interfaces can additionally be accommodated through oxygen off-stoichiometry \cite{Lee2011giant}, in-plane lattice rotation \cite{Zhang2021controlling}, or interface reconstruction \cite{Cheng2018interface}, as shown in Fig. \ref{fig:RMOinterface}. It is possible that coherently strained epitaxial films could be achieved in hexagonal $R$MnO$_3$ films if grown on substrates with smaller lattice mismatch. This, however, would require the design and development of new substrates that offer a better compatibility with the family of hexagonal $AB$O$_3$ materials.

\begin{figure}
    \centering
    \includegraphics[width=\columnwidth]{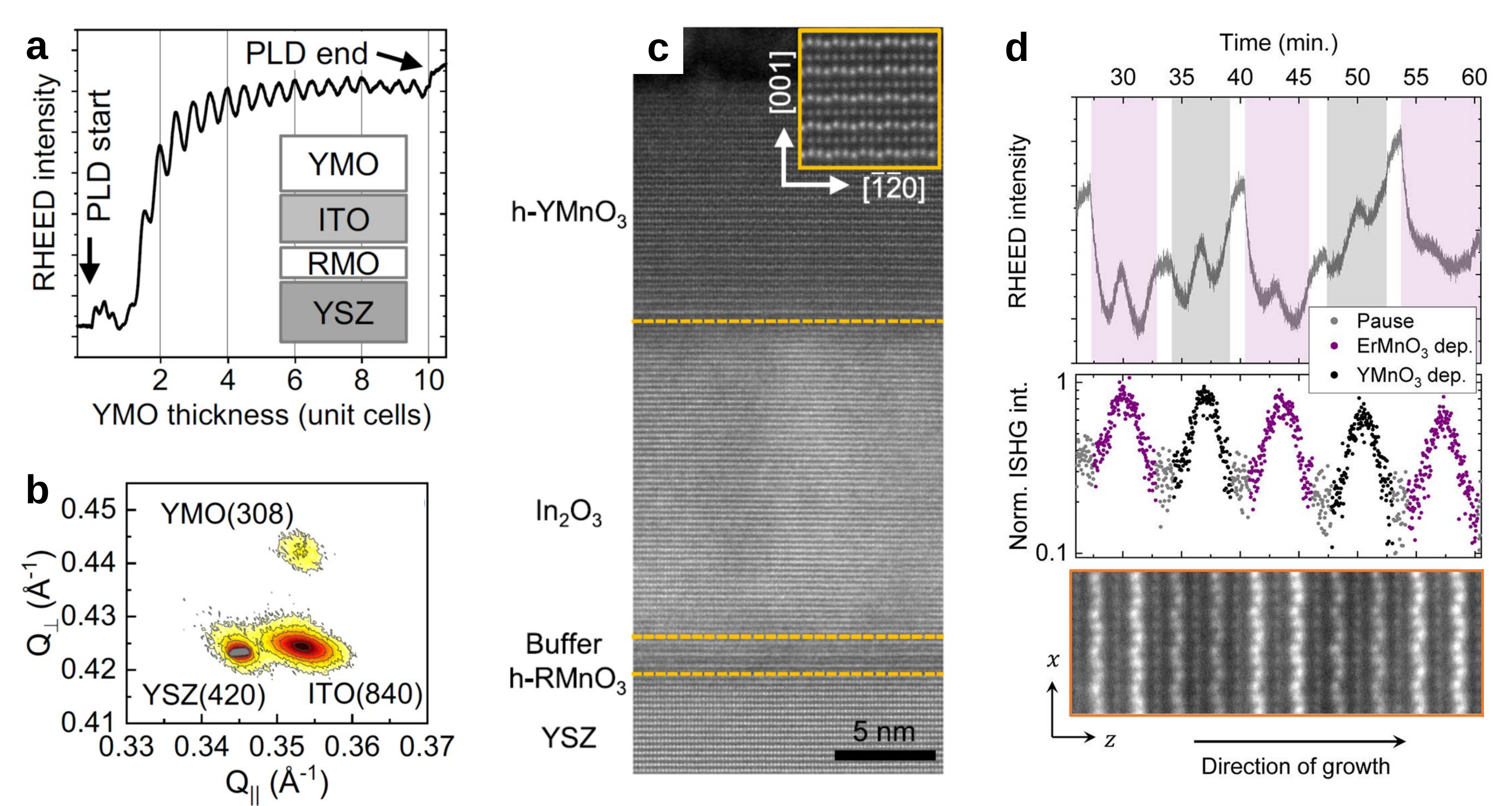}
    \caption{Hexagonal $R$MnO$_3$ in epitaxial heterostructures. (a)-(c) Mutual lattice matching is achieved between YMnO$_3$ and indium tin oxide (SnO$_2$:In$_2$O$_3$, ITO) by inserting a buffer hexagonal $R$MnO$_3$ layer at the substrate interface to induce immediate strain relaxation. (a) RHEED oscillations indicating a layer-by-layer growth mode of YMnO$_3$ in this heterostructure. (b) X-ray reciprocal space mapping shows matching of in-plane lattice parameters between ITO and YMnO$_3$, yet relaxed compared to the underlying YSZ substrate. (c) HAADF-STEM image of a similar heterostructure. (a)-(c) Reprinted figures with permission from Nordlander \textit{et al}. Phys. Rev. Lett. \textbf{4}, 124403 (2020). Copyright (2020) by the American Physical Society.\cite{Nordlander2020epitaxial} (d) A (YMnO$_3$)$_1$/(ErMnO$_3$)$_1$ superlattice grown on YSZ(111). In-situ optical second harmonic generation reveals the inversion-symmetry breaking of each half-unit-cell layer. Reprinted with permission from Nordlander \textit{et al}. Nano Lett. \textbf{21}, 2780-2785 (2021). Copyright (2021) American Chemical Society.\cite{Nordlander2021inversion}}
    \label{fig:RMOheterostructure}
\end{figure}

An alternative route to achieve coherently strained heterostructures is demonstrated through the recent realization of mutual lattice matching between $R$MnO$_3$ and In$_2$O$_3$-based transparent conducting layers such as indium-tin oxide (ITO). Straining the conducting layer to the $R$MnO$_3$ lattice \cite{Nordlander2020epitaxial}, rather than vice versa, offers a new opportunity for epitaxial integration of hexagonal $AB$O$_3$ into functional oxide-electronic heterostructures and superlattices (Fig. \ref{fig:RMOheterostructure}(a)-(c)).

The layered structure of hexagonal $AB$O$_3$ further distinguishes this class of oxides as prospective quantum materials. As already seen in Sec. \ref{sec:intro}, the triangular sub-lattice of each half-unit-cell layer breaks inversion symmetry (see Fig. \ref{fig:structure}(b)). Given the half-unit-cell layer-by-layer growth mode achieved by PLD \cite{Nordlander2020epitaxial}, the symmetry of the ultrathin thin-film system can be alternatingly controlled between preserved and broken inversion symmetry, purely based on either an even or odd total number of half-unit-cell layers \cite{Nordlander2021inversion}. This effect was demonstrated using in-situ SHG \cite{Nordlander2018Probing,Sarott2021Insitu} during hexagonal $R$MnO$_3$ epitaxial growth, where the emission of frequency-doubled light was turned on and off in the films by the deposition of each monolayer (Fig. \ref{fig:RMOheterostructure}(d)). Similar to 2D-layered van-der-Waals materials \cite{Kumar2013second}, such a layer-dependent symmetry control in thin films could furthermore be used to not only tune nonlinear optical responses, but also may present an avenue to engineer, e.g., magnetoelectric, nonreciprocal, chiral or topological effects in the general class of hexagonal $AB$O$_3$ and their heterostructures.

Beyond the hexagonal manganites, the polar hexagonal $P$6$_{\mathrm{3}}cm$ phase has been achieved as bulk crystals for certain gallates ($A$GaO$_3$, $A$ = Y, Er, Ho \cite{Geller1975crystal}) and indates ($A$InO$_3$, $A$ = Y, Sm-Ho, \cite{Pistorius1976stability,Shukla2018rare}) and the non-polar $P$6$_{\mathrm{3}}/mmc$ hexagonal structure is the bulk-stable phase of InFeO$_3$\cite{Nodari1988caracterisation,Giaquinta1994structure}. Although thin films of hexagonal gallates and indates remain little explored so far, thin films of InFeO$_3$ have been grown by PLD on both ZnO(0001) and Ta:SnO$_2$-buffered Al$_2$O$_3$ substrates \cite{seki2010epitaxial,zhang2020infeo3}. InFeO$_3$ films have been proposed as candidates for watersplitting and photoelectrode applications due to their beneficial band-gap placement \cite{zhang2020infeo3}. Additionally, epitaxial InFeO$_3$ has been used as structurally compatible non-polar spacers in hexagonal LuFeO$_3$-based superlattices grown by oxide MBE \cite{holtz2021dimensionality}. The high volatility of indium and its sub-oxides \cite{Vogt2015competing} at the elevated temperatures required for crystallization of the hexagonal phase, however, renders the thin-film synthesis of the indium-based hexagonal compounds challenging; such challenges can also be expected for hexagonal gallates given the similar volatility of gallium.

We note, however, that epitaxy of indates and gallates offers interesting opportunities to study the thin-film functional properties of hexagonal oxides in absence of transition metal ions and magnetism on the $B$ site. This enables, for example, isolating the physics of frustrated rare-earth magnetism on the hexagonal $A$-site lattice \cite{clark2019two}. Moreover, in contrast to the transition-metal-based hexagonal $AB$O$_3$ oxides, which are mixed electronic and ionic conductors, the indates and gallates are electronically insulating and could thus support pure oxygen-ion conductivity.

\section{\label{sec:metastable} Epitaxy of metastable hexagonal $AB$O$_3$ compounds}

Thin film epitaxy is also a powerful platform for exploring metastable hexagonal oxides.  During crystal growth, the phase that is deposited minimizes the global energy of the system. The Gibbs free energy of formation for a crystal nucleating on a substrate is comprised of a volume term, which will prefer the thermodynamically stable structure; a surface term, which is determined by the surface energy between the material and the substrate; and a stress term, which discourages the growth of highly stressed states. In the early stages of growth, the crystal nucleating on the substrate has a high surface-to-volume ratio, making the interface between the substrate and nucleating crystal a critical component in film growth.  Because coherent crystalline interfaces require less energy to form than non-coherent ones, the film often adopts a structure which resembles the crystallographic structure of the substrate in epitaxial thin film growth.  One of the powerful consequences of this minimization of interfacial energy between the substrate and film -- known as epitaxial stabilization -- is that a metastable crystal structure, not the lowest energy structure, can be grown \cite{gorbenko2002epitaxial, dankov1939laws}. By careful choice of substrate\cite{hamasaki2016crystal, katayama2017epitaxial}, epitaxial stabilization enables the exploration of compounds in their metastable and non-equilibrium structures, opening the door to novel materials with new functionalities.  

In the case of rare-earth manganites, $R$MnO$_3$, the ground-state structure depends largely on the $R$-site radius, as discussed in Sec. \ref{sec:structure}. As the radius of the $R$-cation is decreased, the lower energy structure goes from orthorhombic (La – Dy) to hexagonal (Ho – Lu). The small energy difference in the bulk stable state near the crossover between orthorhombic and hexagonal (Fig. \ref{fig:freeenergy}) can be overcome by epitaxial stabilization. For example, Bosak et al. demonstrated that EuMnO$_3$, GdMnO$_3$, and DyMnO$_3$ could form the hexagonal phase by epitaxial stabilization on YSZ (111)\cite{bosak2002epitaxial}, and that HoMnO$_3$, TmMnO$_3$ and LuMnO$_3$ could form the orthorhombic phase using  LaAlO$_3$ (001) or SrTiO$_3$ (001) substrates\cite{bosak2001epitaxial} (Table \ref{tab:stability}).  

Notably, unlike the manganites, there are no known bulk-stable hexagonal ferrites besides InFeO$_3$\cite{Nodari1988caracterisation,Giaquinta1994structure}. Instead, the $R$FeO$_3$ ferrites are usually orthorhombic in their ground state, due to the crystal field splitting of the orbitals as discussed in section \ref{sec:structure}. However, many of the $R$FeO$_3$ structures have been epitaxially stabilized for $R$ with smaller cation radius (Y, Sc, Eu, Tb – Lu), as summarized in Table \ref{tab:stability}. In addition to epitaxial stabilization, ``stromataxic stabilization" can be used to further increase the number of compounds which can be synthesized in the thin film form\cite{garten2021stromataxic}.  Garten et al. showed that by using sequential atomic layering they could form hexagonal ScFeO$_3$ with the YMnO$_3$-like structure. In the same growth conditions, co-deposition of scandium and iron yielded the bixbyite polymorph. Stromataxic stabilization is a powerful technique that could likely be used to construct other metastable hexagonal oxides, beyond those which can be constructed by epitaxial stabilization alone.

\begin{center}
\begin{table}
 \begin{tabular}{||c|c c|c c c|c c|c c c||} 
 \hline
 & \multicolumn{5}{c |}{Mn} & \multicolumn{5}{c ||}{Fe} \\
 \hline
 & \multicolumn{2}{c|}{Stable Structure} & \multicolumn{3}{c|}{Epitaxially Stabilized} & \multicolumn{2}{c|}{Stable Structure} & \multicolumn{3}{c||}{Epitaxially Stabilized} \\
 \hline
 La & o\cite{moussa1996spin,kamata1979thermogravimetric} & Pbnm &  &  &  & o\cite{geller1956crystallographic} & Pbnm & & & \\
 Ce & o\cite{Madelung2000} & Pbnm &  &  &  & o\cite{robbins1969preparation} & Pnma & & & \\
 Pr & o\cite{dabrowski2005structural} & Pbnm &  &  &  & o\cite{geller1956crystallographic} & Pbnm & & & \\
 Nd & o\cite{dabrowski2005structural,mccarthy1973crystal} & Pbnm & & & & o\cite{geller1956crystallographic} & Pbnm & & & \\
 Sm & o\cite{kamata1979thermogravimetric,mccarthy1973crystal} & Pnma & & & & o\cite{geller1956crystallographic} & Pbnm & & & \\
 Eu & o\cite{dabrowski2005structural,mccarthy1973crystal} & Pbnm & h\cite{bosak2002epitaxial} & P6$_3$cm & YSZ(111) & o\cite{geller1956crystallographic} & Pbnm & h\cite{bossak2004xrd} &  P6$_3$cm & YSZ(111) \\
 Gd & o\cite{mccarthy1973crystal} & Pnma & h\cite{bosak2002epitaxial} & P6$_3$cm & YSZ(111) & o\cite{geller1956crystallographic} & Pbnm & & & \\
 Tb & o\cite{quezel1977magnetic} & Pnma & & & & o\cite{marezio1970crystal} & Pbnm & h\cite{akbashev2011weak} & P6$_3$cm & YSZ(111) \\
 Dy & o\cite{kamata1979thermogravimetric,dabrowski2005structural,mccarthy1973crystal} & Pbnm & h\cite{bosak2002epitaxial} & P6$_3$cm & YSZ(111) & o\cite{marezio1970crystal} & Pbnm & h\cite{akbashev2011weak} & P6$_3$cm & YSZ(111) \\
 Ho & h\cite{mccarthy1973crystal, yakel1963crystal} & P6$_3$cm & o\cite{bosak2001epitaxial} & Pnma & LAO(001) & o\cite{marezio1970crystal} & Pbnm & h\cite{akbashev2011weak} & P6$_3$cm & YSZ(111) \\
 & & & & & STO(001) & & & & & \\
 Er & h\cite{kamata1979thermogravimetric, yakel1963crystal, chueh2013intercalation} & P6$_3$cm & & & & o\cite{marezio1970crystal} & Pbnm & h\cite{bossak2004xrd} & P6$_3$cm & YSZ(111) \\
 Tm & h\cite{mccarthy1973crystal, yakel1963crystal} & P6$_3$cm & o\cite{bosak2001epitaxial} & Pnma & LAO(001) & o\cite{marezio1970crystal} & Pbnm & h\cite{bossak2004xrd} & P6$_3$cm & YSZ(111) \\
 & & & & & STO(001) & & & & & \\
 Yb & h\cite{kamata1979thermogravimetric,mccarthy1973crystal, yakel1963crystal} & P6$_3$cm & & & & o\cite{marezio1970crystal} & Pbnm & h\cite{bossak2004xrd} & P6$_3$cm & YSZ(111) \\
 Lu & h\cite{mccarthy1973crystal, yakel1963crystal} & P6$_3$cm & o\cite{bosak2001epitaxial} & Pnma & LAO(001) & o\cite{marezio1970crystal} & Pbnm & h\cite{bossak2004xrd, akbashev2012reconstruction} & P6$_3$cm & YSZ(111) \\
 & & & & & STO(001) & & & & & Al$_2$O$_3$(0001) \\
 & & & & & & & & & & Fe$_3$O$_4$ (111)\\
 Sc & h\cite{greedan1995synthesis} & P6$_3$cm & & & & b\cite{breard2011investigation} & Ia$\overline{3}$ & h\cite{hamasaki2016crystal} & P6$_3$cm & Al$_2$O$_3$(0001) \\
 In & h\cite{bekheet2016ferroelectric} & P6$_3$cm & & & & h\cite{Giaquinta1994structure} & P$6_3/mmc$ & & & \\
 Y & h\cite{kamata1979thermogravimetric, yakel1963crystal} & P6$_3$cm & o & Pnma & LAO(001) & o\cite{geller1956crystallographic} & Pbnm & h\cite{Ahn2013artificially}& P6$_3$cm & Pt(111) \\
 & & & & & STO(001) & & & & & \\
 \hline
\end{tabular}

\caption{Summary of the $R$MnO$_3$ and $R$FeO$_3$ structures where $R$ = rare earth, Sc, In, Y.  The bulk stable phase is listed in addition to the structures which can be epitaxially stabilized on the listed substrate or template.  o, h, b refer to orthorhombic, hexagonal and bixbyite respectively. }
\label{tab:stability} 
\end{table}
\end{center}

\begin{figure}
    \centering
    \includegraphics{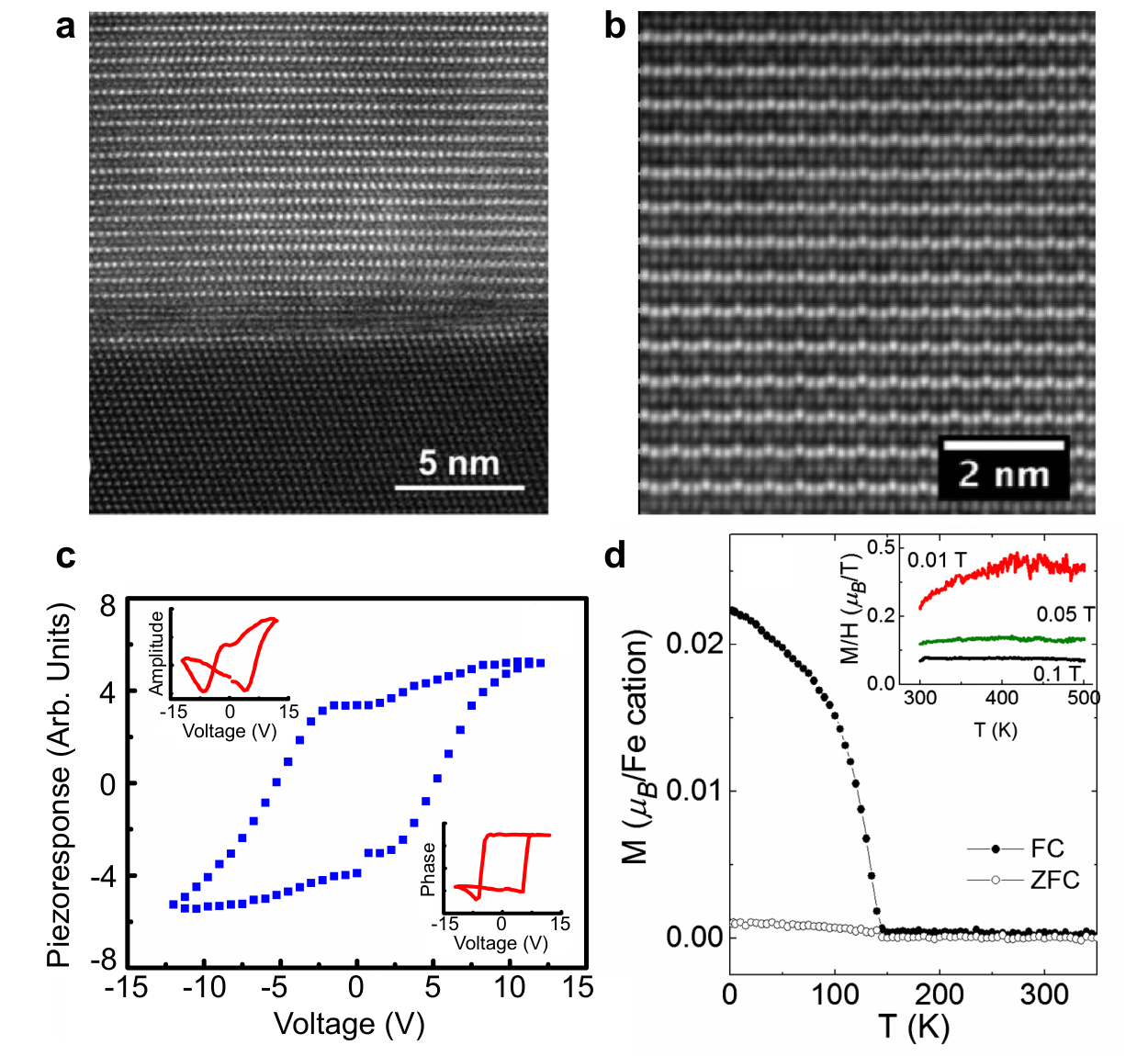}
    \caption{Characterization of epitaxially-stabilized LuFeO$_3$ thin films. (a) HAADF STEM micrograph of a LuFeO$_3$ thin film on YSZ (111). Republished with permission from Akbashev \textit{et al}. CrystEngComm. \textbf{14}, 5373-5376 (2012); permission conveyed through Copyright Clearance Center, Inc. Copyright 2012 The Royal Society of Chemistry.\cite{akbashev2012reconstructed}. (b) HAADF STEM micrograph showing the trimerization of the brighter lutetium atoms characteristic of the P6$_3$cm structure. (c) PFM measurement of LuFeO$_3$/Pt/Al$_2$O$_3$ showing a hysteresis loop with phase and amplitude curves inset. Reprinted figure with permission from Wang \textit{et al}. Phys. Rev. Lett. \textbf{110}, 237601 (2013). Copyright (2013) by the American Physical Society.\cite{wang2013room} (d) Magnetization vs. temperature in an 0.01 T field applied out of plane of a 250 nm film of LuFeO$_3$ on YSZ (111) indicating an weak AFM transition at 143 K with inset high temperature behavior at various applied fields. (b) and (d) Reprinted with permission from Disseler \textit{et al}. Phys. Rev. Lett. \textbf{114}, 217602 (2015). Copyright (2015) by the American Physical Society.\cite{disseler2015magnetic}}
    \label{fig:LuFeO3composite}
\end{figure}

Stabilizing these hexagonal $A$FeO$_3$ ferrites opens the door to further engineering their ferroelectric, magnetic, and multiferroic properties. Here we will focus on LuFeO$_3$, since it has seen a recent surge in interest – for an overview of the other $A$FeO$_3$ compositions, refer to an earlier review of hexagonal ferrites \cite{xu2014multiferroic}.  While LuFeO$_3$ is stable in the orthorhombic structure in the bulk, it can be epitaxially stabilized to be isostructural to YMnO$_3$  \cite{bossak2004xrd}, as shown in Table \ref{tab:stability}. In addition to YSZ \cite{bossak2004xrd, moyer2014intrinsic}, LuFeO$_3$ has been epitaxially stabilized on the basal plane of Al$_2$O$_3$ \cite{disseler2015magnetic}, Ir(111), Pt(111)  \cite{jeong2012epitaxially}, Fe$_3$O$_4$(111)  \cite{akbashev2012reconstruction}, Sn:In$_2$O$_3$  \cite{barrozo2020defect},  and GaN  \cite{casamento2020multiferroic}.  Figure \ref{fig:LuFeO3composite}a,b show high-resolution HAADF-STEM images of epitaxially stabilized LuFeO$_3$ synthesized by MOCVD (Fig. \ref{fig:LuFeO3composite}(a)) and MBE (Fig. \ref{fig:LuFeO3composite}(b)). This hexagonal structure is an improper ferroelectric with P6$_3$cm symmetry (similar to what was discussed in the previous section \ref{sec:bulkstable}) with a polarization of 6.5 $\mu$C/cm$^2$  \cite{jeong2012epitaxially} and a Curie Temperature around 1020 K  \cite{disseler2015magnetic}. The switching of the polarization by piezoforce microscopy\cite{wang2013room} is shown in Fig. \ref{fig:LuFeO3composite}(c).  The hexagonal symmetry of the structure produces a frustrated system, that results in two-dimensional arrangements of spins in the iron planes, as shown in Fig. \ref{fig:magnetism}, resulting in an antiferromagnet. Below approximately 145 K, the spins cant in the [0001] direction, producing a net magnetic moment, displayed in Fig. \ref{fig:LuFeO3composite}(d)\cite{moyer2014intrinsic, disseler2015magnetic}.

\begin{figure}
    \centering
    \includegraphics[width=\textwidth]{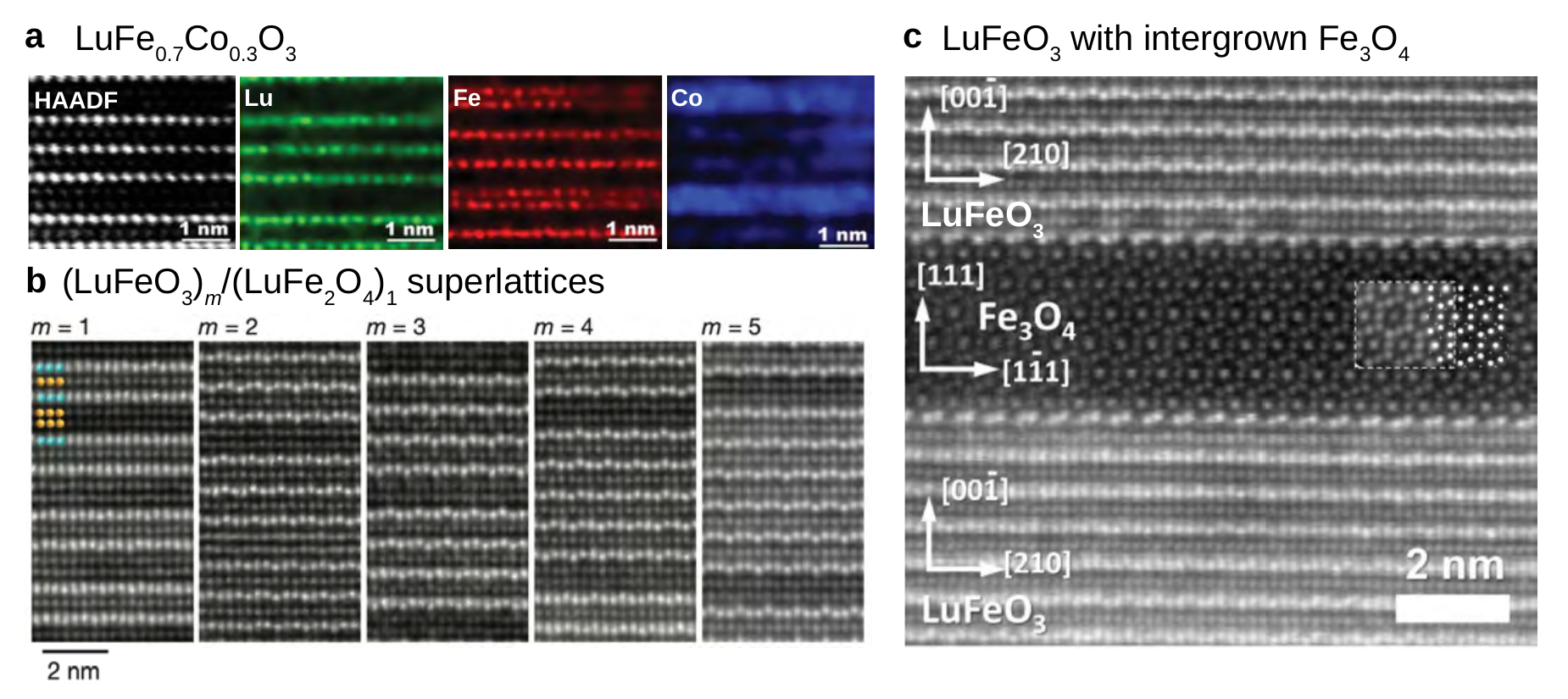}
    \caption{STEM micrographs of LuFeO$_3$ thin films combined with other materials. (a) HAADF STEM and atomic resolution EDS of a small region of cobalt-doped LuFeO$_3$, nominally LuFe$_{0.7}$Co$_{0.3}$O$_3$. Republished with permission from Akbashev \textit{et al}. CrystEngComm. \textbf{14}, 5373-5376 (2012); permission conveyed through Copyright Clearance Center, Inc. Copyright 2012 The Royal Society of Chemistry.\cite{akbashev2012reconstructed}. (b) HAADF-STEM micrographs of the $m$ = 1-5 members of the (LuFeO$_3$)$_m$/(LuFe$_2$O$_4$)$_1$ superlattice series imaged along the LuFeO$_3$ [100] zone axis. The leftmost image shows the position of lutetium (teal) and iron (yellow) atomic columns. Reprinted by permission from Springer Nature Customer Service Centre GmbH: Mundy \textit{et al}. Nature \textbf{537}, 523-527 (2016). Copyright 2016 Springer Nature.\cite{mundy2016atomically} (c) HAADF-STEM micrograph of an iron-rich LuFeO$_3$ film with ingrown epitaxial Fe$_3$O$_4$ nanolayers. Inset shows a HAADF-STEM simulation of Fe$_3$O$_4$ along with white dots that mark the imaged iron positions. Reprinted by permission from Springer Nature Customer Service Centre GmbH: Springer Nature; Akbashev, A \textit{et al}. Sci. Rep. \textbf{2}, 672 (2012). Copyright 2012 The Authors.\cite{akbashev2012reconstruction}}
    \label{fig:LuFeO3wPhases}
\end{figure}

\begin{figure}
	\centering
	\includegraphics[width=\columnwidth]{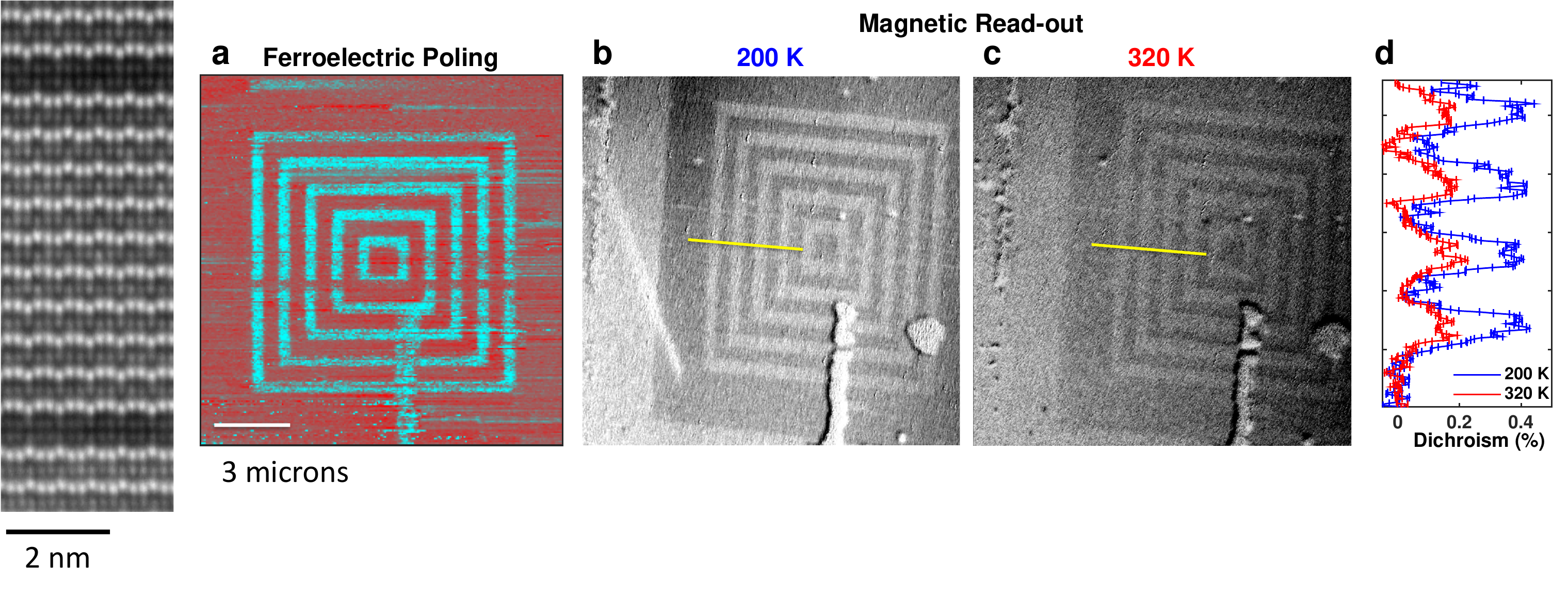}
	\caption{Magneto-electric coupling in (LuFeO$_3$)$_m$/LuFe$_2$O$_4$ superlattices. Left, a HAADF STEM micrograph of a (LuFeO$_3$)$_9$/LuFe$_2$O$_4$ superlattice showing one period of the layered structure. (a) an out-of-plane PFM image at 300\,K of an electrically poled region of the $m=9$ superlattice film showing a pattern of up (teal) and down (red) $c$-axis polarized domains. (b) An XMCD PEEM image of the Fe L$_3$ edge of the poled region in (a) taken at 200\,K. (c) The same image as (b) collected at 320\,K. (d) Line profiles of the XMCD signal along the yellow lines in (b) and (c). The correspondance between the PFM and PEEM images demonstrates the coupling between improper ferroelectricity and ferrimagnetism in the film. Reprinted by permission from Springer Nature Customer Service Centre GmbH: Mundy \textit{et al}. Nature \textbf{537}, 523-527 (2016). Copyright 2016 Springer Nature.\cite{mundy2016atomically}}
	\label{fig:LFO}
\end{figure}

Due to the simultaneous existence of ferroelectricity and a weak ferromagnetic moment, LuFeO$_3$ is a promising candidate for multiferroic materials. In particular, compared to the hexagonal manganites, the hexagonal ferrites promise higher magnetic transition temperatures and magnetoelectric coupling due to the additional occupied orbital \cite{das2014bulk} (the five $d$ electrons in Fe$^{+3}$ half fill each of the orbitals in Fig. \ref{fig:CEFsplitting}).  While LuFeO$_3$ itself only has a net magnetic moment at cryogenic temperatures, LuFeO$_3$ can be synthesized in a superlattice with other materials in an effort to enhance the magnetism.  The first attempts included cobalt doped LuFeO$_3$\cite{akbashev2012reconstructed}, depicted in Fig. \ref{fig:LuFeO3wPhases}(a).  It was was found that these materials showed a decrease in the magnetic ordering temperatures with respect to LuFeO$_3$ and LuFe$_{2-x}$Co$_x$O$_4$.  More recently, LuFeO$_3$ was grown in a superlattice structure with formula-unit-thick layers of LuFe$_2$O$_4$, a hexagonal ferrite with a 240\,K ferrimagnetic transition temperature, to create a room-temperature multiferroic  \cite{mundy2016atomically}. In these superlattices shown in Fig. \ref{fig:LuFeO3wPhases}(b), the ferroelectric rumpling from the LuFeO$_3$ reduced the magnetic frustration in the LuFe$_2$O$_4$, increasing the magnetic transition temperature to 281\,K. The material further demonstrated electric field control of magnetism -- creating a magnetoelectric multiferroic (Fig. \ref{fig:LFO}). This opens doors to the new engineering of multiferroic materials based on these epitaxially stabilized, improper ferroelectrics. We further note that LuFeO$_3$ can be epitaxially stabilized on intergrown spinel Fe$_3$O$_4$ (magnetite)\cite{akbashev2012reconstruction}.  The very high magnetic ordering temperature of Fe$_3$O$_4$ and other spinel compounds could provide a pathway to create a multiferroic with simultaneous transitions well above room-temperature. 

In addition to their coupled electrical and magnetic order, hexagonal ferrites, just as the hexagonal manganites, host topological defects such as their ferroelectric domain walls and vortices where six ferroelectric domains come to a point. The superlattice construction of the (LuFeO$_3$)$_m$ / LuFe$_2$O$_4$ provides an avenue to manipulate these topological textures in these hexagonal ferroelectrics \cite{holtz2021dimensionality}. As the confinement of the ferroelectric layer is increased, the material goes from stabilizing charged domain walls with 3 and 5-fold fractional vortices, to having irregular domains with neutral domain walls. As the confinement approaches one monolayer of LuFeO$_3$, the ferroelectricity is suppressed – reminiscent of the ferroelectric to paraelectric transition with temperature. 

\section{\label{sec:outlook} Outlook}
The hexagonal $AB$O$3$ materials are a rich class of quantum materials.  In addition to the well-studied multiferroic in the hexagonal manganites, there has been recent interest in stabilizing the elusive quantum spin liquid\cite{clark2019two} and predicted topological semimetal states\cite{weber2019topological} and unconventional superconductivity\cite{hu2015predicting}.  Thin film epitaxy provides a powerful platform for both studying these materials at the fundamental limits\cite{nordlander2019ultrathin} as well as potentially stabilizing new materials.  More recent work has used precise chemical layering\cite{garten2021stromataxic} to access compounds that were not accessible with epitaxial stabilization alone. Combined thin film epitaxy should allow additional emergent phenomena to be uncovered in the hexagonal oxides. 

\begin{acknowledgments}
\noindent The authors would like to thank Hena Das and EliseAnne Koskelo for fruitful discussions. This work is supported by the Air Force Research Laboratory, Project Grant FA95502110429. J.A.M acknowledges support from the Packard Foundation and Gordon and Betty Moore Foundation’s EPiQS Initiative, Grant GBMF6760. J.N. acknowledges support from the Swiss National Science Foundation under Project No. P2EZP2\textunderscore195686.
\end{acknowledgments}

\bibliography{biblio}% Produces the bibliography via BibTeX.

\section{A Note on Hexagonal Geometry}
The effective hexagonal in-plane lattice parameter of a (111)-oriented cubic crystal depends on the type of cubic structure the material forms. The simplest structure is the simple cube. For a crystal with cations on the corners of a cube, the (111) lattice parameter is given by the length of the face diagonal of the cube: 
$$
a_{hex} = a_{SC}\sqrt{2}
$$
Materials that follow this formula include cubic perovskites (e.g. SrTiO$_3$, KTaO$_3$, and LaAlO$_3$). Most of the common substrates and bottom electrode materials considered in this review have a face-centered cubic (FCC) structure. For this structure, the hexagonal lattice parameter is given by (1/2) of the face diagonal of the cubic unit cell: 
$$
a_{hex} = \frac{a_{FCC}\sqrt{2}}{2}
$$ 
Various single elements (e.g. Ir, Pt, Pd, Ag, Si) crystallize in an FCC lattice. Spinel (e.g. MgAl$_2$O$_4$) and rock salt-type structures (e.g. MgO) also follow the FCC formula. Other materials with an FCC-type configuration include SiC(3C), CaF$_2$, and Yttria-stabilized Zirconia (YSZ). Some elements crystallize as a body-centered cubic (BCC) lattice (e.g. Cr, Mn, Fe, Mo). The hexagonal lattice parameter for this structure is the length of the face diagonal, just as in the simple cubic case, because the central atom lies slightly out of the (111) plane that includes 3 corner atoms. For the more complicated bixbyite structure, the true (111) lattice parameter follows the simple cubic formula ($a_{hex} = a_{bixbyite}\sqrt{2}$). However, there are cations in a slightly distorted hexagonal lattice with average spacing $a_{hex}/4$. For this reason, the hexagonal lattice parameter of the bixbyite compound In$_2$O$_3$ is reported as 3.58 \AA \ above while the cubic lattice parameter is 10.12 \AA, giving 14.31 \AA \  between identical lattice sites in the (111) plane. Other bixbyite materials include $R_2$O$_3$, Y$_2$O$_3$, and Mn$_2$O$_3$.

Lattice matching between hexagonal crystals is more complicated than that of cubic or orthorhombic structures. While both structures can accommodate matching between crystals with integer ratios between their parameters (e.g. a and 2a), hexagonal films can also incorporate scaling factors like $\frac{3}{2}$ or $\frac{4}{3}$ as well as a 30$^\circ$ rotation between the film and substrate. The rotation effectively scales the lattice parameter by $\sqrt{3}$. This possible rotation, along with a scarcity of substrates with proper lattice parameters, complicates studies of strain engineering in hexagonal thin films. As with cubic thin films, the lattice mismatch between substrate and film is quantified as 
$$
\text{lattice mismatch} \ (\%)= \frac{a_{sub}-a_{film}}{a_{film}} \times 100
$$
For many materials, this simple formula can accurately predict which substrates and films are compatible and how they will align. However, this calculation is imperfect. Especially for large lattice mismatch, the difference in lattice parameters does not tell the whole story. The exact termination of the substrate and structure of the sublattice -- the positions and identities of atoms that lie within the cations that define the hexagonal unit cell -- can change how the substrate and film will align to reduce energy at the interface. Furthermore the tendency of hexagonal crystals to adopt a 30$^\circ$ rotation or unusual scaling factor between film and substrate complicates the quantification of the lattice mismatch because the alignment can dramatically affect the effective lattice parameter of the substrate.

Below we tabulate the lattice parameters used for substrate and film materials (Table \ref{tab:substratelattice} and Table \ref{tab:filmlattice}, respectively) shown in Fig. \ref{fig:numberline2d}.

\begin{center}
\begin{table}
 \begin{tabular}{||c|c|c|c||} 
 \hline
 Compound & Orientation & Hexagonal Lattice Constant (\AA) & 30$^\circ$ Rotated Lattice Constant (\AA) \\
 \hline
 $\alpha$Al$_2$O$_3$ & (0001) & 4.76 & 8.24 \\
 YSZ & (111) & 3.64 & 6.30 \\
 GaN & (0001) & 3.19 & 5.52 \\
 SCAM & (0001) & 3.25 & 5.62 \\
 4H-SiC & (0001) & 3.07 & 5.32 \\
 ZnO & (0001) & 3.25 & 5.63 \\
 Graphite & (0001) & 2.46 & 4.26 \\
 In$_2$O$_3$ & (111) & 14.31 (3.58) & 24.79 (6.20) \\
 CaF$_2$ & (111) & 3.86 & 6.69 \\
 KTaO$_3$ & (111) & 5.64 & 9.77 \\
 LaAlO$_3$ & (111) & 5.37 & 9.30 \\
 MgAl$_2$O$_4$ & (111) & 5.71 & 9.90 \\
 CoFe$_2$O$_4$ & (111) & 3.67 & 6.36 \\
 MgO & (111) & 2.98 & 5.16 \\
 3C-SiC & (111) & 3.08 & 5.33 \\
 SrTiO$_3$ & (111) & 5.52 & 9.56 \\
 Pt & (111) & 2.78 & 4.82 \\
 Pd & (111) & 2.75 & 4.76 \\
 Ir & (111) & 2.72 & 4.70 \\
 Si & (111) & 3.84 & 6.65 \\
 \hline
 \end{tabular}
 \caption{Equivalent hexagonal lattice parameters for the substrate materials considered in Fig. \ref{fig:numberline2d}.}
 \label{tab:substratelattice}
 \end{table}

\begin{table} 
 \begin{tabular}{||c|c|c|c|c||}
 \hline
 Film & Geometry & Lattice Constant (\AA) & 30$^\circ$ Rotated Lattice Constant (\AA) & Reference \\
 \hline
 YMnO$_3$ & bulk & 6.13 & & \cite{yakel1963crystal} \\
 GdMnO$_3$ & thin film & 6.30 & & \cite{lee2007epitaxial} \\
 HoMnO$_3$ & bulk & 6.14 & & \cite{yakel1963crystal}\\
 ErMnO$_3$ & bulk & 6.12 & & \cite{yakel1963crystal} \\
 TmMnO$_3$ & bulk & 6.06 & & \cite{yakel1963crystal} \\
 YbMnO$_3$ & bulk & 6.06 & & \cite{yakel1963crystal} \\
 LuMnO$_3$ & bulk & 6.04 & & \cite{yakel1963crystal} \\
 TbMnO$_3$ & thin film & 6.27 & & \cite{lee2006epitaxial} \\
 DyMnO$_3$ & bulk & 6.19 & & \cite{harikrishnan2009phase} \\
 LuFeO$_3$ & bulk & 5.97 & & \cite{bossak2004xrd, akbashev2012reconstruction} \\
 LuFe$_2$O$_4$ & bulk & 3.43 & 5.96 & \cite{iida1990single} \\
 DyFeO$_3$ & thin film & 6.24 & & \cite{kasahara2021room} \\
 DyFe$_2$O$_4$ & thin film & 3.54 & 6.13 & \cite{steinhardt2021dyfe2o4} \\
 YbFeO$_3$ & thin film & 3.46 & 5.99 & \cite{iida2012ferroelectricity} \\
 ErFeO$_3$ & thin film & 6.05-6.09 & & \cite{yokota2015examination, bossak2004xrd} \\
 ScFeO$_3$ & thin film & 5.72 & & \cite{hamasaki2020switchable} \\
 TmFeO$_3$ & thin film & 6.02 & & \cite{bossak2004xrd} \\
 YFeO$_3$ & bulk & 3.51 & 6.08 & \cite{li2008hexagonal} \\
 InFeO$_3$ & thin film & 3.32 & 5.75 & \cite{seki2010epitaxial} \\
 InFe$_2$O$_4$ & thin film & 3.36 & 5.82 & \cite{seki2010epitaxial} \\
 InMnO$_3$ & bulk & 5.88 & & \cite{belik2009magnetic} \\
 YInO$_3$ & bulk & 6.27 & & \cite{paul2016evolution} \\
 HoInO$_3$ & bulk & 6.27 & & \cite{paul2016evolution} \\
 DyInO$_3$ & bulk & 6.30 & & \cite{paul2016evolution} \\
 TbInO$_3$ & bulk & 6.32 & & \cite{paul2016evolution} \\
 GdInO$_3$ & bulk & 6.35 & & \cite{paul2016evolution} \\
 EuInO$_3$ & bulk & 6.38 & & \cite{paul2016evolution} \\
 SmInO$_3$ & bulk & 6.42 & & \cite{paul2016evolution} \\
 YGaO$_3$ & bulk & 6.07 & & \cite{adem2007ferroelectric} \\
 InGaO$_3$ & bulk & 3.31 & 5.73 & \cite{belik2009magnetic} \\
 \hline
\end{tabular}
\caption{Lattice parameters of known hexagonal $AB$O$_3$ compounds.}
\label{tab:filmlattice}
\end{table}
\end{center}
\end{document}